\documentclass[fleqn,usenatbib]{mnras}

\usepackage{newtxtext,newtxmath}

\usepackage[T1]{fontenc}
\usepackage{ae,aecompl}

\usepackage{graphicx}	
\usepackage{amsmath}	
\usepackage{xcolor}
\usepackage{longtable}
\usepackage{lscape}

\title[{\it Kepler} detached EBs: radii and photometric masses]{Detached eclipsing binaries from the {\it Kepler} field: radii and photometric masses of components in short-period systems}

\author[P. Cruz et al.]{
Patricia Cruz$^{1,2}$\thanks{E-mail: pcruz@cab.inta-csic.es (PC); marcos.diaz@iag.usp.br (MD)},
John F. Aguilar$^{3}$,
Hern\'an E. Garrido$^{4}$,
Marcos P. Diaz$^{5}$, and
\newauthor{Enrique Solano$^{1,2}$}
\\
$^{1}$Depto. de Astrof{\'i}sica, Centro de Astrobiolog{\'i}a (INTA-CSIC), ESAC campus, Camino Bajo del Castillo s/n, E-28692, Villanueva de la Ca{\~n}ada, Spain\\
$^{2}$Spanish Virtual Observatory (SVO), E-28692, Villanueva de la Ca{\~n}ada, Spain\\
$^{3}$Departamento de Matem\'aticas, Universidad Militar Nueva Granada, kil\'ometro 2 v\'ia Cajic\'a - Zipaquir\'a, Colombia, c\'odigo postal 110111.\\
$^{4}$Universidad de C\'ordoba, Departamento de F\'isica y Electr\'onica, Cra.\ 6A Nº 77-305, Monter\'ia, C\'ordoba, Colombia\\
$^{5}$Instituto de Astronomia, Geof\'isica  e Ci{\^e}ncias Atmosf\'ericas, Universidade de S{\~a}o Paulo, Rua do Mat{\~a}o 1226, Cidade Universit\'aria, 05508-090 S{\~a}o Paulo, Brazil\\
}

\date{Accepted 2022 June 15. Received 2022 June 15; in original form 2022 February 15}

\pubyear{2022}

\begin{document}
\label{firstpage}
\pagerange{\pageref{firstpage}--\pageref{lastpage}}
\maketitle

\begin{abstract}
 The characterisation of detached eclipsing binaries with low mass components has become important when verifying the role of convection in stellar evolutionary models, which requires model-independent measurements of stellar parameters with great precision. However, spectroscopic characterisation depends on single-target radial velocity observations and only a few tens of well-studied low-mass systems have been diagnosed in this way. We characterise eclipsing detached systems from the {\it Kepler} field with low mass components by adopting a purely-photometric method. Based on an extensive multi-colour dataset, we derive effective temperatures and photometric masses of individual components using clustering techniques. We also estimate the stellar radii from additional modelling of the available {\it Kepler} light curves. Our measurements confirm the presence of an inflation trend in the mass-radius diagram against theoretical stellar models in the low-mass regime.
\end{abstract}




\section{Introduction}

The characterisation of eclipsing binaries (EBs), especially detached systems with low mass components (with $M_{\star} \leq 0.7 M_{\odot}$), has become a good approach for testing stellar evolutionary models, concerning the role of convection in these later type stars \citep[][]{Feiden12,Han19}. This requires unbiased, high-precision measurements of stellar masses and radii in large homogeneous samples. 
By combining spectroscopic and photometric time-series data, it is possible to determine the physical properties (radii and masses) of EB components with uncertainties of $5\%$ or less \citep[e.g.][]{Torres2002,Lopez-Morales2005,Birkby2012}. Such well characterised detached systems in very close orbits -- with orbital periods ($P_{\rm orb}$) of 2-3 days or less -- show discrepancies when compared to stellar models: the estimated radii can be $5$-to-$20\%$ larger than predicted \citep[][]{Lopez-Morales2005,Kraus2011,Cruz2018,Chaturvedi2018}. However, just a few tens of well-charaterised detached EB (DEB) systems are available in the literature \citep[][]{Southworth2015,Chaturvedi2018}.

In order to explain the measured anomalous radius of low-mass stars, several scenarios were proposed in the past years and remain discussed to date in the literature. For example, \citet{Lopez-Morales2007} proposed that the stellar metallicity would play a role, as it was observed that isolated metal-rich stars presented larger radii than expected. The magnetic activity was also pointed as responsible, where M-dwarf stars in detached short-period systems would have an enhanced activity, which would inhibit convection and cause their radius to inflate \citep{Chabrier2007,Kraus2011}. Moreover, part of the radius anomaly problem could come from the current stellar models, which are not able to fully reproduce the properties of active low-mass star \citep{Morales2010,Irwin2011}.

It is important to increase the sample of known low-mass DEBs to investigate the radius anomaly causes. Nevertheless, the proper spectroscopic characterisation is time-consuming and depends on a lot of telescope time. \citet[][]{Garrido2019} have then adopted a purely-photometric method to characterise $230$ short-period DEBs from the Catalina Sky Survey \citep[CSS][]{Drake2009}, with $P_{\rm orb} < 2$ days, using available broad-band photometric data and CSS light curves (LCs) only. The mentioned method provided the fractional radius of each component, estimated from light-curve modeling. They also derived photometric masses, which were obtained based on a multi-color dataset by using clustering techniques, as a confirmation of the low-mass nature of the binary components \citep[for more details, see][]{Garrido2019}. Regardless large individual uncertainties, their work have considerably increased the number of known systems with main-sequence low-mass components.

The use of machine learning algorithms has become more frequent and has been adopted for data mining and automated classification in astronomical databases \citep[][and references therein]{Sanchez2013,Chattopadhyay2014}. 
In this work, we aim to identify new DEB systems with low mass components. For that, we adopted unsupervised and supervised methods to classify previously identified binary systems from the {\it Kepler} space mission \citep[][]{Borucki2010} according to their luminosity class, using a multi-colour dataset, and derive the effective temperature and the photometric mass of each individual component by searching for similarities between observed data and models, as done in \citet[][]{Garrido2019}. Moreover, we used the available {\it Kepler} LCs to derive the fractional radius and orbital parameters of selected EB systems.

This paper is organised as follows. The sample selection is described in Sect. \ref{selection}. The characterisation of each binary component individually is presented in Sect. \ref{charact}, which includes the description of the adopted machine learning methods. The analysis is presented in Sect. \ref{model}, which shows the adopted LC modelling procedure and the obtained results. In Sect. \ref{discuss}, we present the comparison with previous works and a discussion on the obtained mass-radius diagram, and finally in Sect. \ref{concl}, our conclusions.

\section{Sample Selection}\label{selection}

We selected eclipsing binary systems from the Kepler Eclipsing Binary Catalog\footnote{Available at \url{http://keplerebs.villanova.edu}.} \citep[Third Revision,][hereafter KEBC]{Kirk2016}, with orbital period of $4$ days or less, and with effective temperature ($T_{\rm eff}$) up to $6\,000$K. Note that $T_{\rm eff}$ available at the catalogue was derived by the authors from the stellar energy distribution (SED), based on broad-band photometry and assuming a single star configuration \citep[for details, see][]{Kirk2016}. Therefore, these temperatures were used only as a selection criterion.

The selection was cross-matched to broad-band photometric catalogs, keeping only those EB systems with available data from the Panoramic Survey Telescope and Rapid Response System \citep[Pan-STARRS;][]{Tonry2012} and Two-Micron All-Sky Survey \citep[2MASS;][]{Skrutskie2006}. Only those with data in all offered filters -- $g\,r\,i\,z\,y$ from Pan-STARRS and $J\,H\,K_{\rm s}$ from 2MASS -- were kept. Differently from our previous work \citep[][]{Garrido2019}, we used the Pan-STARRS photometry to cover the region of the visible because the Sloan Digital Sky Survey \citep[SDSS;][]{Abazajian2009} is not available for the whole {\it Kepler} field. Finally, we also excluded those systems that presented any warning flag in the available broad-band photometry, indicating possible spurious measurements. We selected, then, a list of $821$ EB systems to undergo a light curve (LC) inspection.

\subsection{Light curves}\label{LCs}

\begin{figure}
\centering
	\includegraphics[width=0.98\columnwidth]{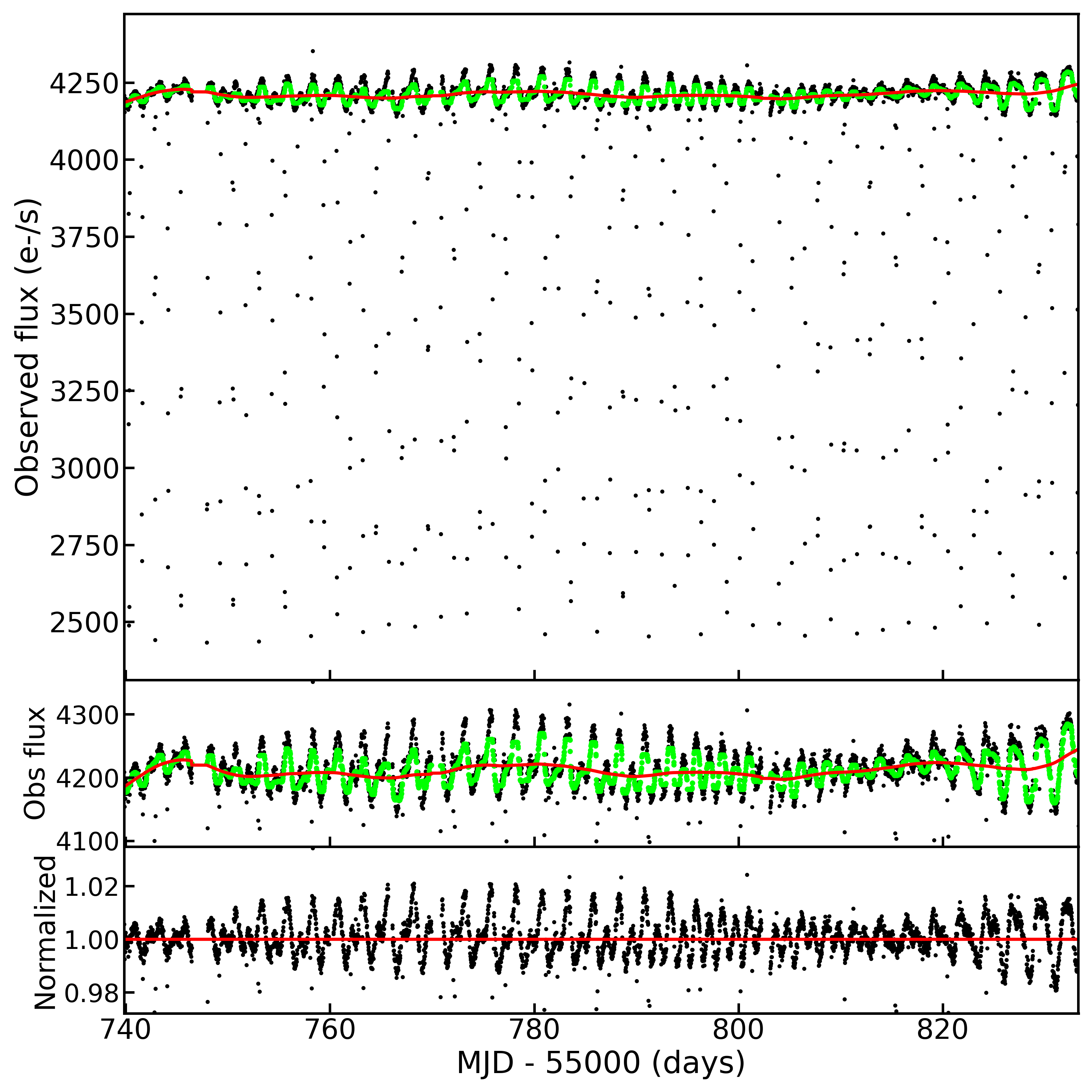}
	\caption{Normalization of the tenth quarter of observations for EB KIC09656543. {\sl Top panel:} The black dots show {\it Kepler} light curve, with 4179 epochs. The green curve is the calculated median curve, and the red line shows the spline fit. {\sl Middle panel:} The same as previous plot, but zoomed in the LC baseline. {\sl Bottom panel:} The obtained light curve, showing the resulting normalization along the baseline.}
    \label{fig2_Q4normalization}
\end{figure}

The light curves were obtained from {\it Kepler}'s database, where we downloaded the long-cadence data from all observed quarters for each selected EB. 
To obtain normalized LCs, we adopted the {\it sliding median} method, dealing with each available quarter of observations separately for each object individually. A median LC is generated by sliding a window with a fixed size (in number of epochs) along the light curve and replacing the central point by the calculated median within the window. The size of the window is different for each quarter and each object, as it is calculated as being the one with the highest chi-square when both LCs are compared, the original and the generated median LC, after iteration using different window sizes. The median LC was then fitted by a cubic spline function to obtain a more smooth normalization curve, which was used to correct the original light curve. 
Figure \ref{fig2_Q4normalization} shows the results of such normalization procedure for KIC09656543 as an example. Top and middle panels show the calculated median light curve in green, resulting from the sliding median window, and the red curve represents the normalization function obtained by the spline fit. The normalized light curve (for tenth quarter) is presented on the bottom panel. 
Repeating the same procedure with all available quarters for a given EB, we obtained the complete normalized light curve.

\subsection{Sample of detached binaries}

\citet{Matijevic2012} adopted an automatic procedure, the Locally Linear Embedding (LLE) method, to derive the morphological class of eclipsing binaries -- distinguishing between detached, semi-detached, and contact systems -- which is presented as a morphology value\footnote{The morphology value, {\sl morph} or {\sl c}, is a classification parameter that varies between $0$ and $1$.}, {\sl c}, in the KEBC catalogue \citep[][]{Kirk2016}. According to \citet{Matijevic2012}, if the classification parameter $c$ is less than $0.5$, a system is probably a detached system. If $c$ lays between $0.5$ and $0.7$, a system is more likely to be a semi-detached binary and the contact systems may have $c$ between $0.7$ and $0.8$. These authors mentioned that well-detached system should have $c < 0.1$ and almost sinusoidal LCs should have $c > 0.9$. They also stated that the derived morphology value presented in the catalog should be taken only as a guideline, as the separation between different classes is not abrupt. For instance, there are detached systems with estimated $c\sim 0.7$, and semi-detached ones with $c\sim 0.4$ \citep[for example, see][Fig. 4]{Matijevic2012}.

Therefore, all light curves underwent an independent visual inspection to separate them into the three morphological classes. For that, we generated the phase-folded curves by adopting the binary orbital period given in the KEBC catalog \citep{Kirk2016}. Some of the LCs presented intense baseline variation -- what could due to a strong stellar activity or other unidentified cause that turned the visual classification ambiguous -- and, therefore, they were excluded. 
From the $821$ LCs inspected (the selected objects described previously in this section), $300$ of them could be classified as detached EB systems (DEBs). 
As a comparison to the results obtained from the LLE method, which are available at the KEBC catalog, $60$\% of our DEB sample have $c<0.5$ and $48$\% of them have $0.5<c<0.7$.

\section{Characterisation of binary components}\label{charact}

We constructed a ten-colour grid of models with synthetic composite colours -- based on the available broad-band photometric data from Pan-STARRS ($g\,r\,i\,z\,y$) and 2MASS ($J\,H\,K_{\rm s}$) -- to identify the EB systems with only main-sequence stars as components. 
We adopted colour indices to discriminate dwarf stars from systems with evolved components. For that, we used models for giant stars to represent and discriminate from possible evolved stars – like subgiants, for instance – as their spectral energy distribution – and, therefore, their colours – resemble those of giant stars.

This calibration grid was generated based on evolutionary models with $1$, $3$, and $5$ Gyr from \citet{Bressan2012}\footnote{These models are available at \url{http://stev.oapd.inaf.it/cgi-bin/cmd}.}, considering three possible EB configurations: systems composed by two dwarf stars (V+V), by a dwarf and a giant (V+III), and by two giants (III+III), where giant models represent evolved components. 
We used a total of $1\,020$ different models: $903$ models for giant stars, with $T_{\rm eff}$ from $1\,624$ to $5\,979$ K, and $117$ for dwarfs, with $T_{\rm eff}$ from $2\,303$ to $5\,991$ K. These models were combined two-by-two to generate synthetic model binaries, as done by \citet{Parihar2009}, which resulted in a detailed synthetic grid with $180\,579$ different binary models.

For the PanSTARRS-2MASS ten-colour model grid, we adopted the seven standard colours with adjacent passbands -- $(g-r)$, $(r-i)$, $(i-z)$, $(z-y)$, $(y-J)$, $(J-H)$, $(H-K_{\rm s})$ -- added to the same three redder colours adopted by \citet{Garrido2019} to better identify the V+V binaries -- $(r-J)$, $(r-H)$ and $(z-K_{\rm s})$.

\subsection{Defining temperatures using machine learning algorithms}\label{KNN}

We searched for similarities between the observed photometry and models to classify our DEB sample according to their luminosity class, and derive the effective temperature of each individual component. For that, we followed the same methodology described in \citet{Garrido2019}, which is briefly described below.

We adopted the $k$-means clustering technique, starting with the Hartigan test \citep{Hartigan1975}, which was applied to the calibration grid with over $180\,000$ synthetic binaries jointly with the $821$ selected EBs (see Sect. \ref{selection}), aiming at optimising the number of clusters that should be formed. The Hartigan test reached the optimal solution of $17$ clusters after 50 iterations. We then performed the $k$-means method to assign the data to each cluster, by distributing the whole dataset (models and observed data) into $17$ separate clusters. The cluster centroids are initially randomly defined, and then recalculated over several iterations until they reach convergence, where distances between the centroid and the data are minimal\footnote{For more details on this clustering procedure, see Sect. 3 of our previous work \citep{Garrido2019}.}.

\begin{table*}
\centering
\scriptsize
\caption{Effective temperatures and photometric masses obtained for a test set of well known V+V binaries, with available Pan-STARRS and 2MASS photometry. The estimated $T_{\rm eff}$ and $M_{\star}$ values are shown in columns 2-5 and the comparison values (from literature) are presented in columns 6-9. The respective references are presented in the last column.}
\begin{tabular}{l|cccc|cccc|c}
\hline\hline	
Star Name    &	$\mathrm{T}_1$ (K)          	&	 $\mathrm{T}_2$ (K)        &  $\mathrm{M}_1$ (M$_{\odot}$) &  $\mathrm{M}_2$ (M$_{\odot}$)   &   $\mathrm{T}_{1,pub}$ (K)       &	 $\mathrm{T}_{2,pub}$ (K)	      &	 $\mathrm{M}_{1,pub}$ (M$_{\odot}$)		    &		$\mathrm{M}_{2,pub}$ (M$_{\odot}$)        & Reference \\   
 & (K) & (K) & (M$_{\odot}$) & (M$_{\odot}$) & (K) & (K) & (M$_{\odot}$) & (M$_{\odot}$) & \\ 
\hline
\multicolumn{9}{l}{\sl Correctly assigned V+V systems}\\
2MASS J01542930+0053266 & 3698$\pm$100 & 2742$\pm$552 & 0.504$\pm$0.032 & 0.091$\pm$0.186 & 3800 & 3600 & 0.515$\pm$0.023 & 0.548$\pm$0.025 & \citet{Becker2008} \\ 
2MASS J10305521+0334265 & 3700$\pm$100 & 3054$\pm$430 & 0.505$\pm$0.032 & 0.161$\pm$0.229 & 3720$\pm$20 & 3630$\pm$20 & 0.499$\pm$0.002 & 0.444$\pm$0.002 & \citet{Kraus2011} \\ 
2MASS J16502074+4639013 & 3337$\pm$100 & 2919$\pm$322 & 0.302$\pm$0.060 & 0.119$\pm$0.129 & 3500 & 3395 & 0.490$\pm$0.003 & 0.486$\pm$0.003 & \citet{Creevey2005} \\ 
2MASS J23143816+0339493 & 3493$\pm$100 & 2935$\pm$338 & 0.396$\pm$0.057 & 0.123$\pm$0.142 & 3460$\pm$180 & 3320$\pm$180 & 0.469$\pm$0.002 & 0.383$\pm$0.001 & \citet{Kraus2011} \\ 
HAT-TR-318-007 & 3180$\pm$100 & 2825$\pm$200 & 0.216$\pm$0.053 & 0.101$\pm$0.050 & 3190$\pm$110 & 3100$\pm$110 & 0.448$\pm$0.011 & 0.2721$\pm$0.0042 & \citet{Hartman2018} \\ 
MOTESS-GNAT 646680 & 3870$\pm$100 & 3526$\pm$153 & 0.545$\pm$0.034 & 0.415$\pm$0.080 & 3730$\pm$20 & 3630$\pm$20 & 0.499$\pm$0.002 & 0.443$\pm$0.002 & \citet{Kraus2011} \\ 
NSVS 11868841 & 5093$\pm$112 & 3531$\pm$815 & 0.804$\pm$0.040 & 0.418$\pm$0.257 & 5250$\pm$135 & 5020$\pm$135 & 0.870$\pm$0.074 & 0.607$\pm$0.053 & \citet{Cakirli2010} \\ 
SDSS-MEB-1 & 3177$\pm$100 & 2984$\pm$121 & 0.215$\pm$0.053 & 0.137$\pm$0.044 & 3320$\pm$130 & 3300$\pm$130 & 0.272$\pm$0.020 & 0.240$\pm$0.022 & \citet{Blake2008} \\ 
V1236 Tau & 5339$\pm$109 & 3800$\pm$895 & 0.899$\pm$0.038 & 0.536$\pm$0.188 & 4200$\pm$200 & 4150$\pm$200 & 0.787$\pm$0.012 & 0.770$\pm$0.012 & \citet{Bayless2006} \\ 
V404 CMa & 4178$\pm$100 & 3159$\pm$490 & 0.641$\pm$0.022 & 0.206$\pm$0.276 & 4200$\pm$100 & 3940$\pm$20 & 0.750$\pm$0.005 & 0.659$\pm$0.005 & \citet{Rozyczka2009} \\ 

WTS 19b-2-01387 & 3617$\pm$100 & 3062$\pm$350 & 0.466$\pm$0.045 & 0.164$\pm$0.183 & 3498$\pm$100 & 3436$\pm$100 & 0.498$\pm$0.019 & 0.481$\pm$0.017 & \citet{Birkby2012} \\ 
WTS 19c-3-01405 & 3597$\pm$119 & 3021$\pm$384 & 0.455$\pm$0.056 & 0.149$\pm$0.193 & 3309$\pm$130 & 3305$\pm$139 & 0.410$\pm$0.023 & 0.376$\pm$0.024 & \citet{Birkby2012} \\ 
WTS 19c-3-08647 & 3963$\pm$100 & 3202$\pm$423 & 0.576$\pm$0.033 & 0.227$\pm$0.242 & 3900$\pm$100 & 3000$\pm$150 & 0.393$\pm$0.019 & 0.244$\pm$0.014 & \citet{Cruz2018} \\ 
WTS 19f-4-05194 & 4370$\pm$117 & 3599$\pm$691 & 0.679$\pm$0.018 & 0.456$\pm$0.209 & 4400$\pm$100 & 3500$\pm$100 & 0.531$\pm$0.016 & 0.385$\pm$0.01 & \citet{Cruz2018} \\ 
WTS 19g-2-08064 & 4443$\pm$100 & 3224$\pm$133 & 0.691$\pm$0.014 & 0.239$\pm$0.075 & 4200$\pm$100 & 3100$\pm$100 & 0.717$\pm$0.027 & 0.644$\pm$0.025 & \citet{Cruz2018} \\ 
WTS 19g-4-02069 & 3186$\pm$100 & 2683$\pm$ 459 & 0.219$\pm$0.053 & 0.087$\pm$0.112 & 3300$\pm$140 & 2950$\pm$140 & 0.53$\pm$0.02 & 0.143$\pm$0.006 & \citet{Nefs2013} \\ 

\multicolumn{9}{l}{\sl Correctly discarded non-V+V systems}\\
AN Cam & 5836$\pm$118 & 5609$\pm$129 & 1.048$\pm$0.047 & 0.970$\pm$0.038 & 6050$\pm$150 & 5750$\pm$150  & 1.380$\pm$0.021 & 1.402$\pm$0.025 & \citet{Southworth2021}\\ 
2MASS J04463285+1901432 & 3545$\pm$100 & 2980$\pm$349 & 0.426$\pm$0.053 & 0.136$\pm$0.162 & 3320$\pm$150 & 2900$\pm$150 & 0.467$\pm$0.050 & 0.192$\pm$0.020 & \citet{Hebb2006} \\ 
LSPM J1112+7626 & 3554$\pm$100 & 2986$\pm$353 & 0.432$\pm$0.053 & 0.138$\pm$0.165 & 3191$\pm$164 & 3079$\pm$166 & 0.395$\pm$0.002 & 0.275$\pm$0.001 & \citet{Irwin2011} \\ 
V1174 Ori & 5299$\pm$562 & 3867$\pm$100 & 0.875$\pm$0.183 & 0.545$\pm$0.033 & 4470$\pm$120 & 3615$\pm$100 & 1.01$\pm$0.015 & 0.731$\pm$0.008 & \citet{Stassun2004} \\ 
\multicolumn{9}{l}{\sl Misclassified V+V systems} \\
2MASSJ03262072+0312362 & 3524$\pm$100 & 2949$\pm$333 & 0.414$\pm$0.055 & 0.127$\pm$0.143 & 3330$\pm$60 & 3270$\pm$60 & 0.527$\pm$0.002 & 0.491$\pm$0.001 & \citet{Kraus2011} \\ 
2MASSJ07431157+0316220 & 3341$\pm$100 & 2909$\pm$309 & 0.305$\pm$0.060 & 0.117$\pm$0.119 & 3730$\pm$90 & 3610$\pm$90 & 0.584$\pm$0.002 & 0.544$\pm$0.002 & \citet{Kraus2011} \\ 
2MASSJ19071662+4639532 & 4651$\pm$385 & 4008$\pm$151 & 0.719$\pm$0.069 & 0.592$\pm$0.044 & 4150 & 3700 & 0.679$\pm$0.010 & 0.523$\pm$0.006 & \citet{Devor2008} \\ 
2MASSJ20115132+0337194 & 5305$\pm$560 & 3819$\pm$100 & 0.879$\pm$0.180 & 0.540$\pm$0.020 & 3690$\pm$80 & 3610$\pm$80 & 0.557$\pm$0.001 & 0.535$\pm$0.001 & \citet{Kraus2011} \\ 

\hline\hline
\end{tabular}
\label{tab:testsample}
\end{table*}

\begin{figure*}
\centering
	\includegraphics[angle=90,width=1.0\columnwidth,trim={2cm 2cm 2cm 3cm},clip]{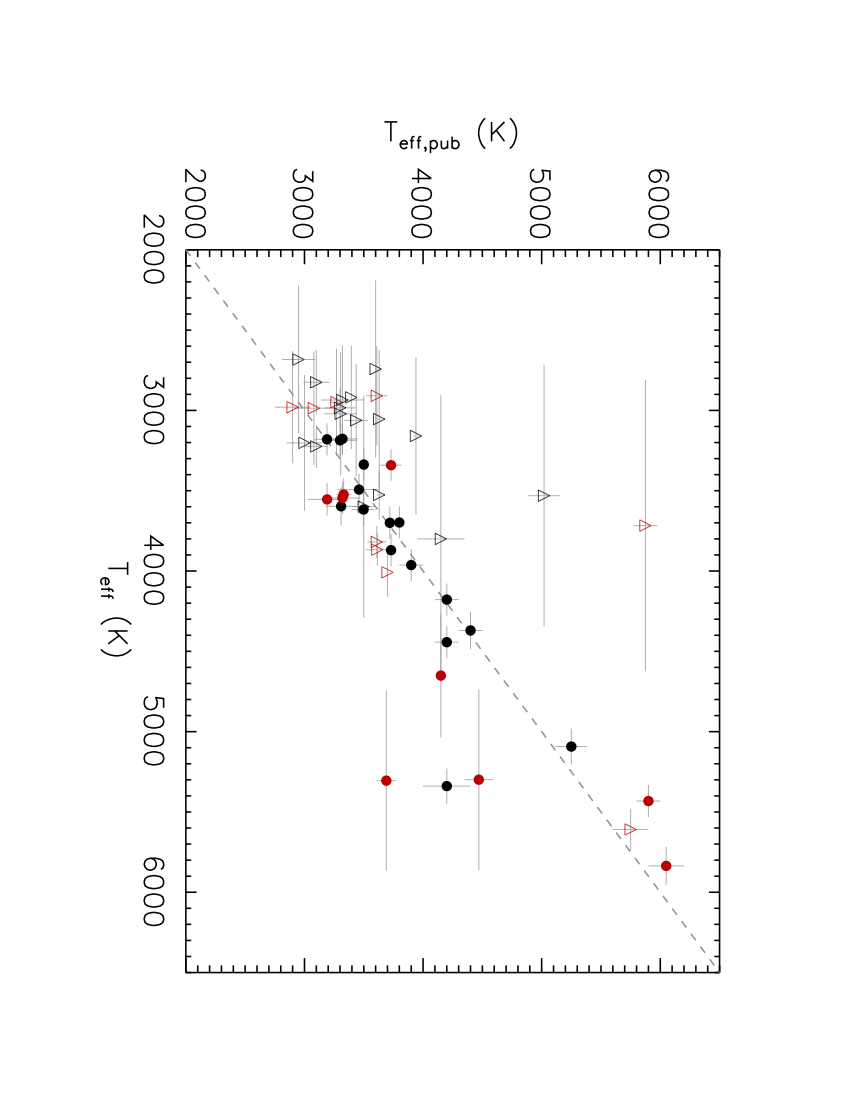}
	\includegraphics[angle=90,width=1.0\columnwidth,trim={2cm 2cm 2cm 3cm},clip]{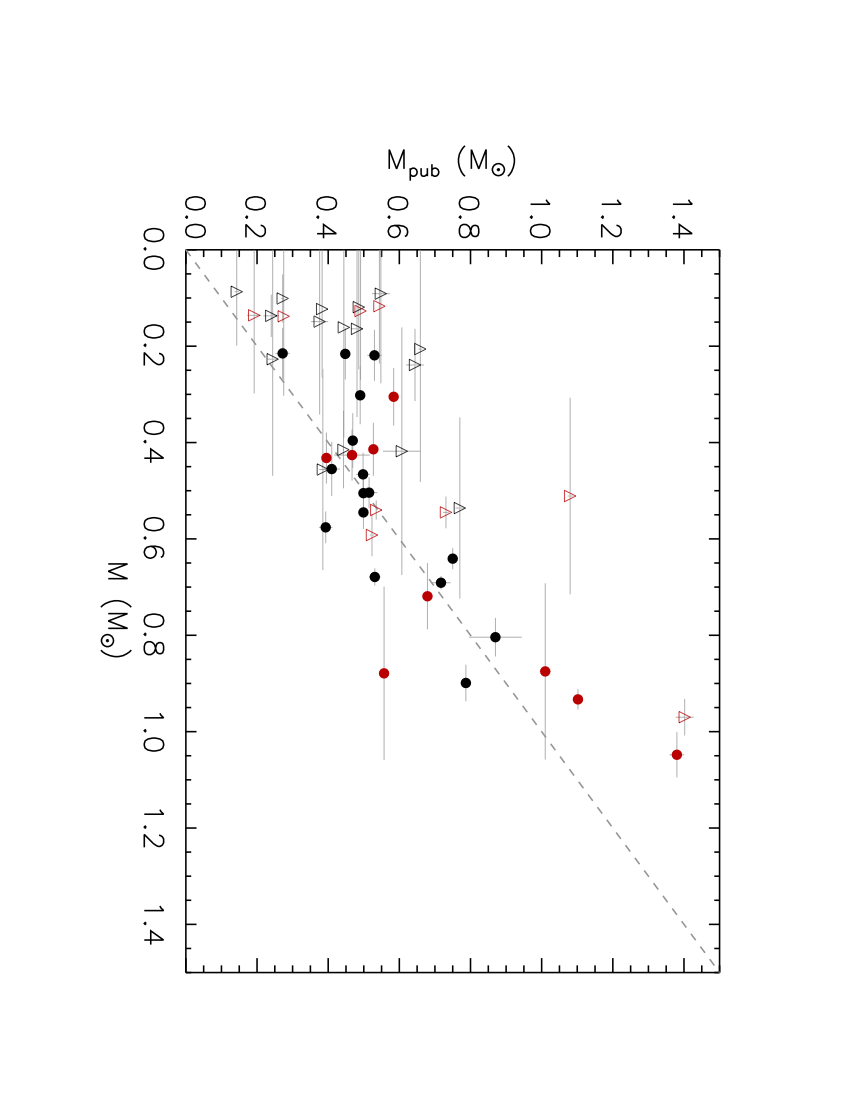}
    \caption{Comparison of obtained effective temperatures (left panel) and masses (right panel) for each EB component of the test sample with the literature values. Primary and secondary components are represented by filled circles and open triangles, respectively. The identified V+V systems are shown in black and the rejected systems are shown in red. The dashed line illustrates the identity function, for reference.}
    \label{fig_controlsample}
\end{figure*}

Once clusters and their members are defined, we then used the K-Nearest Neighbours classifiers \citep[KNN,][]{Cover1967,Chattopadhyay2014} to characterise the systems according to their surrounding models. Considering fixed centroids for the observational data, the nearest synthetic data set were defined using euclidean distances in $\Re^{10}$, taking into account all ten colours of the grid. If sets of models from different EB pairs -- V+V, III+V, and III+III -- were present in a given cluster, the closest set was chosen based on the mean distance between the observational data and all the models in that given set. Hence, the closest set defines the class of our binaries. 
Since each synthetic binary (from models) already has temperatures defined for the primary and the secondary components, the effective temperature of our objects was determined as the mean $T_{\rm eff}$ value of their nearest surrounding models. The uncertainties in temperature were estimated by the standard deviation, as previously done in \citet{Garrido2019}. 

We applied the described procedure to a set of 24 known detached binaries from the literature, where 20 of them are V+V systems. Several of them were selected from the catalogue by \citet[][]{Eker2014}. We selected objects with available Pan-STARSS and 2MASS photometry, with stellar parameters derived from spectroscopic analysis. In general, the estimated effective temperatures are in agreement with literature values, within error estimates, as shown in table \ref{tab:testsample}). The method was able to correctly assign the luminosity class for 16 known V+V systems.

We added to the test sample a binary system with a subgiant component -- AN Cam \citep[][]{Southworth2021} -- aiming at verifying how it would be treated by the method, since the grid of models was constructed with giant and dwarf stellar models only. The method was also successful in rejecting the AN Cam system, as it was identified as V+III, a binary with a giant component. The same way, we also analysed three other EBs, either with a substellar or a subdwarf component: 2MASSJ04463285+1901432 \citep{Hebb2006}, LSPM J1112+7626 \citep[][]{Irwin2011}, and V1174 Ori \citep[][]{Stassun2004}. Although having good $T_{\rm eff}$ estimates, these systems were also discarded by the method, being classified as V+III systems. Although the mentioned systems do not have a giant component, they were correctly discarded as non-V+V systems, being assigned this way probably because the model grid does not have models for substellar nor subdwarf objects included.

The method has misclassified as V+III and, hence, discarded, four known V+V systems, representing $20\%$ of the tested V+V set. We believe that was due to the unrealistic proportion between giant and dwarf models in the grid, as generated V+V systems represent only $\sim$1.3\% of the model binaries. For this work, we used the most complete grid of models available by \citet[][]{Bressan2012} to generate our grid -- combining them in pairs to generate III+III, III+V, and V+V systems -- and this is to be further improved as it misclassified some V+V systems. Nevertheless, this is not critical for our purpose, since we aimed at having a clean sample of V+V EB systems -- i.e, free from giant contaminants -- not focusing on the completeness of the analysed sample. 

To have an estimate of the mass -- hereafter, photometric mass -- we interpolated the obtained temperatures with tabulated values using a forth-order polynomial fit. We adopted the values for effective temperatures and masses of main-sequence stars by \citet[][]{Pecaut13}\footnote{The authors maintain a regular update for the mentioned tabulated values, which are available at \url{https://www.pas.rochester.edu/~emamajek/EEM_dwarf_UBVIJHK_colors_Teff.txt}.}, which is based on semi-empirical values. The photometric mass error estimates were calculated considering the $T_{\rm eff}$ uncertainties obtained with the KNN method. These values are also presented in table \ref{tab:testsample}).

\begin{figure*}
    \centering
    \includegraphics[width=2.1\columnwidth,trim={0cm 0cm 5cm 0cm},clip]{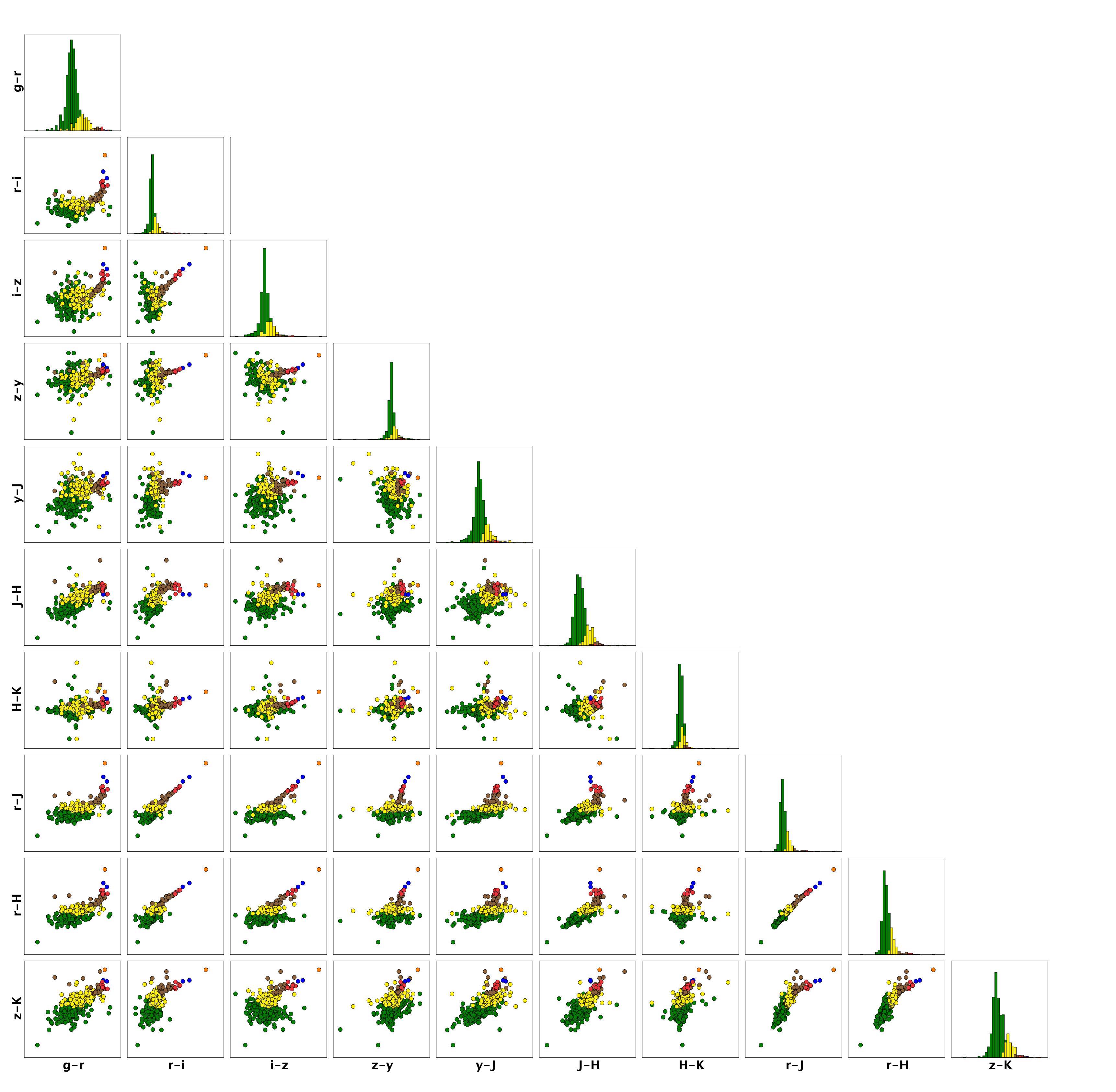}
    \caption{Distribution of the identified $493$ V+V EB systems in the adopted ten-colour grid, shown in colour cuts. The six clusters formed by the $k$-means method in which the V+V systems were distributed are represented here in different colours (as green, yellow, brown, red, blue, and orange bullets).}
    \label{fig9_distibution}
\end{figure*}

Figure \ref{fig_controlsample} shows the comparison of the obtained values for the effective temperature and mass for each component of the test sample with published values (left and right panels, respectively). Primary components are presented as filled circles and secondary components as open triangles. The EBs correctly assigned as V+V systems are shown in black and the rejected systems -- assigned by the method as V+III or III+III systems -- are shown in red. The dashed line represents the identity function. For the estimated temperatures (left panel), a smaller dispersion is seen for the primaries in comparison to the estimated values for the secondaries. This is corroborated by the Pearson correlation coefficient ($r$) of 0.82 and 0.78 for the primary and the secondary components, respectively. It is worth noting that larger $T_{\rm eff}$ uncertainties were estimated of the secondaries. A similar behavior is shown for the estimated photometric masses (fig. \ref{fig_controlsample}, right panel), with $r$ of 0.78 and 0.76 for the primary and the secondary components, respectively. As expected, we obtained larger errors in this case, nevertheless, presenting a good agreement with the values from literature, considering the estimated uncertainties (see table \ref{tab:testsample}). Differences with published values are more notorious for low temperature and mass values -- $T_{\rm eff}<3500$ K and $M_\star< 0.4$ M$_\odot$ -- especially for secondary components. This behaviour was expected, as we used only the information coming from broad-band photometry where a low-mass secondary component would contribute less in the binary integrated flux. Moreover, there could be a limitation due to the adopted colours, since we do not explore the infrared region.

Intrinsic stellar variability, due to starspots for instance, could affect the adopted single-epoch photometric measurements from Pan-STARRS and 2MASS, which could lead to a wrong classification by the method and/or erroneous temperature and mass estimations. Independently of the assigned classification, some temperatures were estimated by the method with large error bars, which could indicate inconsistencies among the photometric data when compared to the models. For instance, some of the outliers in figure \ref{fig_controlsample} are systems with observed stellar variability probably due to starspots. This is the case, for example, of AN Cam \citep{Southworth2021} and V1174 Ori \citep{Stassun2004}, where the observed light curve variability showed strong evidences of spot activity.

We then performed the KNN method for the initial sample with $821$ EB systems, regardless their LC morphology. The model binaries with only dwarf stars as components were present in six of the $17$ clusters formed by the $k$-means method, meaning that only the observed binaries within these clusters could be assigned as V+V systems. 
As a result, we found that $493$ EBs from our selected sample were defined as V+V systems. The remaining systems were discarded by the method, nevertheless, as mentioned previously, this does not mean they have giant components. For example, the EB KIC08736245 system is a previously known system with solar-type components leaving the main sequence \citep[][]{Fetherolf2019}, and it was correctly discarded by our method, as it could not be classified as a V+V system. 
As mentioned previously, the used single-epoch photometric measurements may be affected by intrinsic variability, for example due to starspots, which could lead to an incorrect classification. Therefore, some detached V+V systems may have been discarded by the method. Nevertheless, the main objective of this work was to select a sample clean from evolved components, not aiming at having a complete subset of detached V+V binaries.

Among the identified V+V systems, the obtained $T_{\rm eff}$ vary from $3\,152$ to $5\,991$ K for primary components and from $2\,632$ to $5\,991$ K for the secondaries. Note that the highest estimated temperature ($5\,991$ K) is the upper $T_{\rm eff}$ limit of our model grid. All colour cuts are presented in fig. \ref{fig9_distibution}, showing the distribution of V+V systems in six of the clusters formed by the $k$-means method, where each cluster is represent by a different colour (as green, yellow, brown, red, blue, and orange bullets).

Correlating the V+V EBs found by the KNN method with those systems identified as detached binaries according to their LC morphology (the adopted criteria are described in Sect. \ref{selection}), we gathered a final list with $164$ detached systems with only main-sequence stars. 
The derived photometric masses and effective temperatures for all $330$ stars (both components of the $164$ V+V DEB systems) are presented in tables \ref{Tab:TeffMass1}, \ref{Tab:TeffMass2}, and \ref{Tab:TeffMass3}, in Appendix \ref{ApenTeff}. 
For example, the estimated temperatures for the components of EB KIC12109845 are of $4\,612$ and $3\,971$ K for the primary and secondary components, respectively. This system was previously described as a detached system composed by a late K and an early M stars \citep[][]{Harrison2012}.

\begin{figure}
\centering
	\includegraphics[angle=90,width=1.0\columnwidth,trim={2cm 2cm 3cm 2cm},clip]{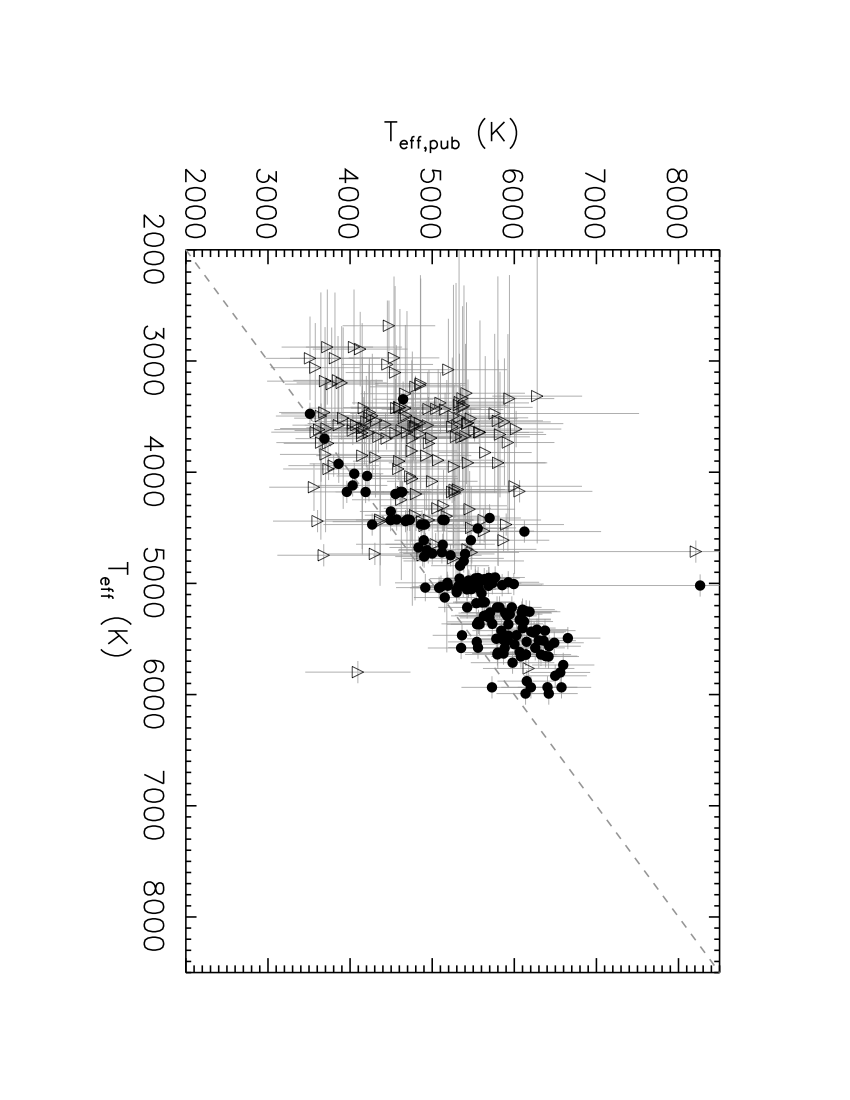}
	\includegraphics[angle=90,width=1.0\columnwidth,trim={2cm 2cm 3cm 2cm},clip]{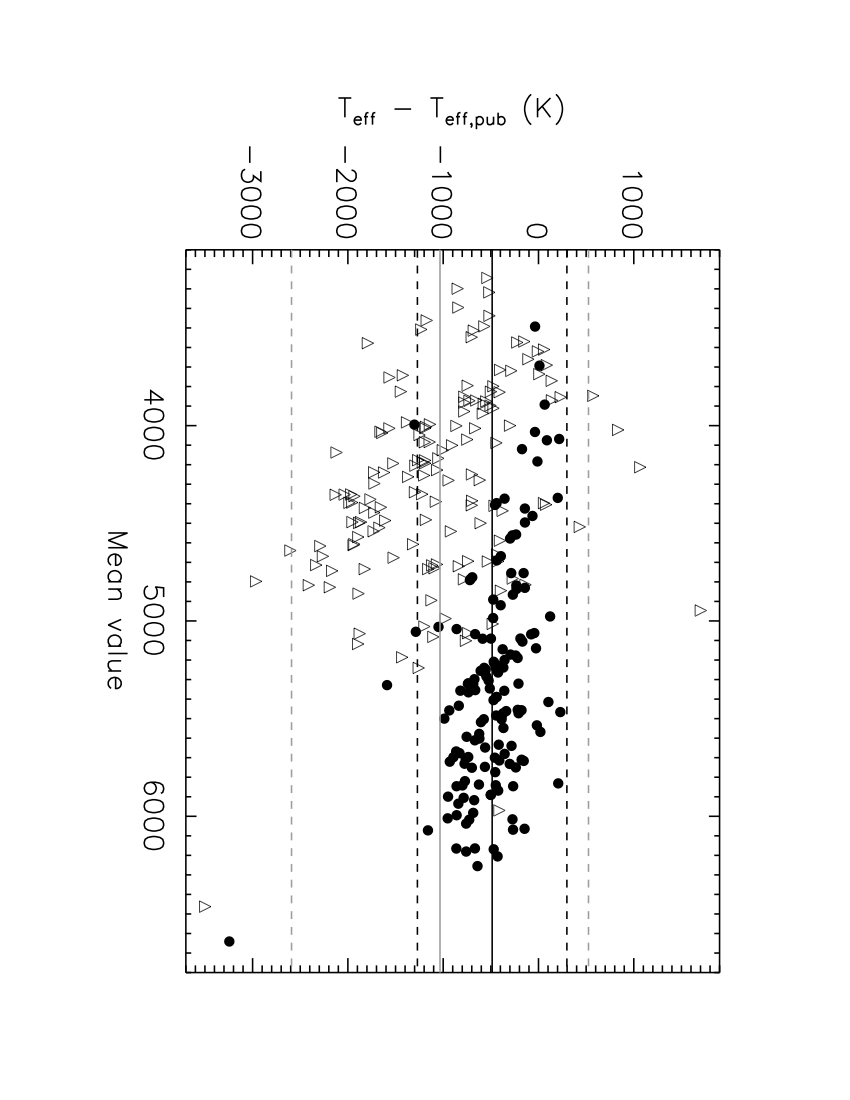}
    \caption{Comparison between effective temperatures obtained in this work with the literature values from \citet{Armstrong2014}. {\sl Top panel:} Our estimated $T_{\rm eff}$ for each {\it Kepler} DEB component are shown in the horizontal axis and the literature values are presented in the vertical axis. Primary and secondary components are represented by filled circles and open triangles, respectively. The dashed line illustrates the identity function. {\sl Bottom panel:} The Bland-Altman plot, showing the agreement between both methods. The horizontal axis shows the mean value between the results from \citet[][]{Armstrong2014} and ours, and the vertical axis shows the difference between them. The mean $T_{\rm eff}$ difference for primaries and secondaries are shown as black and grey solid lines, respectively. Dashed lines illustrate the 2-sigma interval.}
    \label{fig_tempcomp}
\end{figure}

We compared estimated temperatures with the values from \citet{Armstrong2014}. Among the $148$ systems in common to their study, in general these authors estimated higher temperatures for primary and secondary components, as shown in figure \ref{fig_tempcomp} (top panel). As mentioned before, our highest estimated temperature was limited to $5\,991$ K, the hottest dwarf model used in our grid. Some of the comparison temperatures are above $8\,000$ K. 
The agreement between both approaches is also presented (fig. \ref{fig_tempcomp}, bottom panel) in a Bland-Altman plot, illustrating the difference between the results from \citet[][]{Armstrong2014} and ours. The black and grey solid lines show that the mean $T_{\rm eff}$ difference ($T_{\rm eff,this work} - T_{\rm eff,pub}$) is approximately of $-486$ and $-1033$ K for primaries and secondaries, respectively. Dashed lines show the 2-sigma interval for a 95\% confidence level.

\begin{figure}
    \includegraphics[width=0.98\columnwidth]{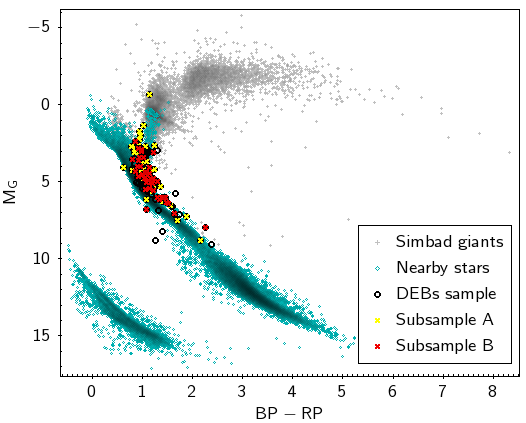}
    \caption{Colour-magnitude diagram of our sample of DEB candidates (black open circles). The overplotted subsamples A and B represent systems with radii values estimated from light-curve fitting (see Sect. \ref{discuss} for details). Nearby main-sequence objects (within $\sim 100$ pc) identified from Gaia EDR3 are shown as cyan diamonds. A sample of SIMBAD giant stars are shown as gray crosses.}
    \label{fig_HRdiag}
\end{figure}

\subsection{Searching for giant contaminants}\label{contaminants}

We searched for giant contaminants within our final sample of $164$ DEBs by comparing our objects to a set of dwarf and giant stars in a colour-magnitude diagram (CMD).

We used the SIMBAD Astronomical Database\footnote{\url{http://simbad.u-strasbg.fr/simbad/}.} \citep{Wenger2000} to find a representative sample of objects classified as giant stars, with spectral types ranging from F to M type. With the help of TOPCAT\footnote{TOPCAT is an interactive {\it Tool for OPerations on Catalogues And Tables}, available at \url{http://www.star.bris.ac.uk/~mbt/topcat/}.} \citep{Taylor2005,Taylor2011}, we searched for the available Gaia parallaxes and photometry by cross-matching the set of giant stars within $5\arcsec$ with the Gaia Early Data Release 3 \citep[Gaia EDR3;][epoch 2016]{Gaia2016,Gaia2021}. 
A second set of comparison was defined from the Gaia EDR3, searching for nearby objects with parallax greater than $10$ mas (within $\sim 100$ pc). From both sets, giant stars from SIMBAD and nearby stars from Gaia, we kept only those objects that presented Gaia parallax error of less than $10$\%, with errors in all Gaia magnitudes ($G$, $BP$, and $RP$ bands) of less than $5$\%, and with the Renormalised Unit Weight Error (RUWE) of less than $1$, which is the value expected for single stars \citep[][]{Arenou2018,Lindegren2018,Lindegren2021}.

To have a clear distribution of comparison stars in the main sequence, we firstly estimated the corrected $BP$ and $RP$ flux excess factor, $C^*$, which is a quality metric according to their $(BP-RP)$ colours, defined by \citet[][]{Riello2021}. In figure \ref{fig_HRdiag}, the greenish diamonds represent the sample of Gaia nearby stars which have $|C^*| < \sigma$, where $\sigma$ is the $C^*$ scatter as function of $G$ magnitude\footnote{For more details on the applied quality criteria, see Table 2, and Eqs. 6 and 18 from \citet[][and references therein]{Riello2021}}. The sample of giant stars from SIMBAD are shown in fig. \ref{fig_HRdiag}, represented by gray crosses.

We also cross-matched our sample ($164$ DEBs) with Gaia EDR3, where we found good photometric data (errors within $5$\%) for $161$ systems. These objects are presented in figure \ref{fig_HRdiag} as black open circles. The overplotted subsamples A and B (shown as yellow and red crosses, respectively) represent those systems with radii estimates from LC modeling, which will be explained later in Section \ref{discuss}. Note that the majority of our EBs have $(BP-RP)$ colour -- combined for the system -- redder than $0.983$, which is the expected value for a K0\,V star according to \citet[][]{Pecaut13}.

\section{Light-curve modelling with JKTEBOP and AGA}\label{model}

We have identified $300$ objects from the {\it Kepler} EBs Catalog -- considering the criteria described in Sect. \ref{selection} -- as short-period detached systems. However, after applying the KNN method (described in Sect. \ref{KNN}), only $164$ DEB candidates were classified as $V+V$ systems. For this sample of $V+V$ DEBs, we performed the fitting of {\it Kepler} LCs using the JKTEBOP code \citep{Southworth2004,Southworth2013}, which uses the Nelson–Davis–Etzel model \citep[NDE model;][]{Nelson1972, Popper1981} and is suitable for detached systems. To help dealing with such long time series -- {\it Kepler} light curves can have tens of thousands epochs -- we adopted the asexual genetic algorithm (AGA) by \citet{Canto2009}, which was implemented to the JKTEBOP code by \citet{Coughlin2011}, and already was successfully applied for hundreds of detached EBs found in the CSS (for more details, see \cite{Garrido2019}).

We calculated the stellar measured flux ($F_{\rm obs}$) from the median flux in each quarter of observations, as shown below:
\begin{equation}
    F_{\rm obs}=\sum_{Q=1}^{n} F_{\rm Q}\frac{N_{\rm Q}}{N_{\rm tot}},
	\label{eq_flux}
\end{equation}
\noindent
where ($F_{\rm Q}$) is the median flux observed in a given quarter, $N_{\rm Q}$ is the number of observations (epochs) in that quarter, $N_{\rm tot}$ is the total number of epochs, and $n$ is the last observed quarter. Note that the maximum value for $n$ is $18$, which is the number of quarters observed during the {\it Kepler} mission. However, not all stars have data for all quarters. We obtained the analysed LCs in magnitudes by adopting the zero point given by the {\it Kepler} team, where the empirical flux for a star with Kepler magnitude of $Kp = 12$ mag is of $1.74\times 10^{5}$ e$^{-}$/s. 

Light curve models provide information on the radius of each component in addition to the orbital configuration of the studied systems. For the modelling, the following parameters were free to vary in the first approach:

\begin{itemize}
\addtolength{\itemindent}{0.7cm}
\item[--] the orbital period ($P_{\rm orb}$),
\item[--] the reference time (epoch) of primary minimum ($T_{\rm 0}$),
\item[--] the sum of the stellar radii ($r_{1}+r_{2}$),
\item[--] the ratio of the radii ($r_{2}/r_{1}$),
\item[--] the central surface brightness ratio ($J=J_{2}/J_{1}$),
\item[--] the orbital inclination ($i$),
\item[--] the orbital eccentricity ($ecc$), and
\item[--] the baseline level of the light curve,
\end{itemize}
\noindent
where $P_{\rm orb}$ and $T_{\rm 0}$ are given in days, $r_{1}+r_{2}$ is given in units of the binary separation, $i$ in degrees, and the baseline level in Kepler magnitudes.

We adopted $P_{\rm orb}$ and $T_{\rm 0}$ values from the KEBC catalogue as initial conditions. 
We assumed that the argument of periastron ($\omega$) to be null, as it may be not well constrained. 
We also assumed no third light source in the binary light curve. The photometric mass ratio ($q$) was calculated from the previously obtained photometric masses (sect. \ref{KNN}) and it was kept fixed, as it is used by the code only to determine possible tidal deformation on the components.

Limb-darkening (LD) coefficients were estimated with the JKTLD\footnote{The JKTLD source code is available at \url{http://www.astro.keele.ac.uk/jkt/codes/jktld.html}.} procedure, considering the temperatures obtained for each component (see sect. \ref{KNN}). We adopted a quadratic LD law with coefficients from \citet{Claret2000}. Both LD coefficients were used as fixed parameters in the modelling procedure with JKTEBOP. Gravity darkening coefficients were also kept fixed to $0.32$, which is the typical value for stars with convective envelopes \citep{Lucy1967}. The reflection coefficients were also fixed in the fitting, calculated by the code based on the system geometry.

The fitting was performed iteratively. 
Some LCs presented non-negligible baseline variation over time -- for example due to spots -- easily identified in the phase-folded light curve containing the complete data set. Since the time-series data can have up to four years of observations, it may affect the eclipse relative depth and the results from light-curve modelling.

Therefore, we adopted a second approach considering only one quarter of observations -- {\it Kepler} observations were divided in $90$-day quarters. We chose the quarter with the highest number of epochs observed for each binary. The fitting with JKTEBOP was performed considering the same variables and assumptions described above, with the exception of $P_{\rm orb}$, which was kept as a fixed value. This is a good assumption too as the orbital period have been already defined with a good precision \citep[][and references therein]{Kirk2016}. This is also a robust approach based on a minimum set of free parameters, which are more reliable within a quarter.

\begin{figure}
    \centering
	\includegraphics[width=0.98\columnwidth]{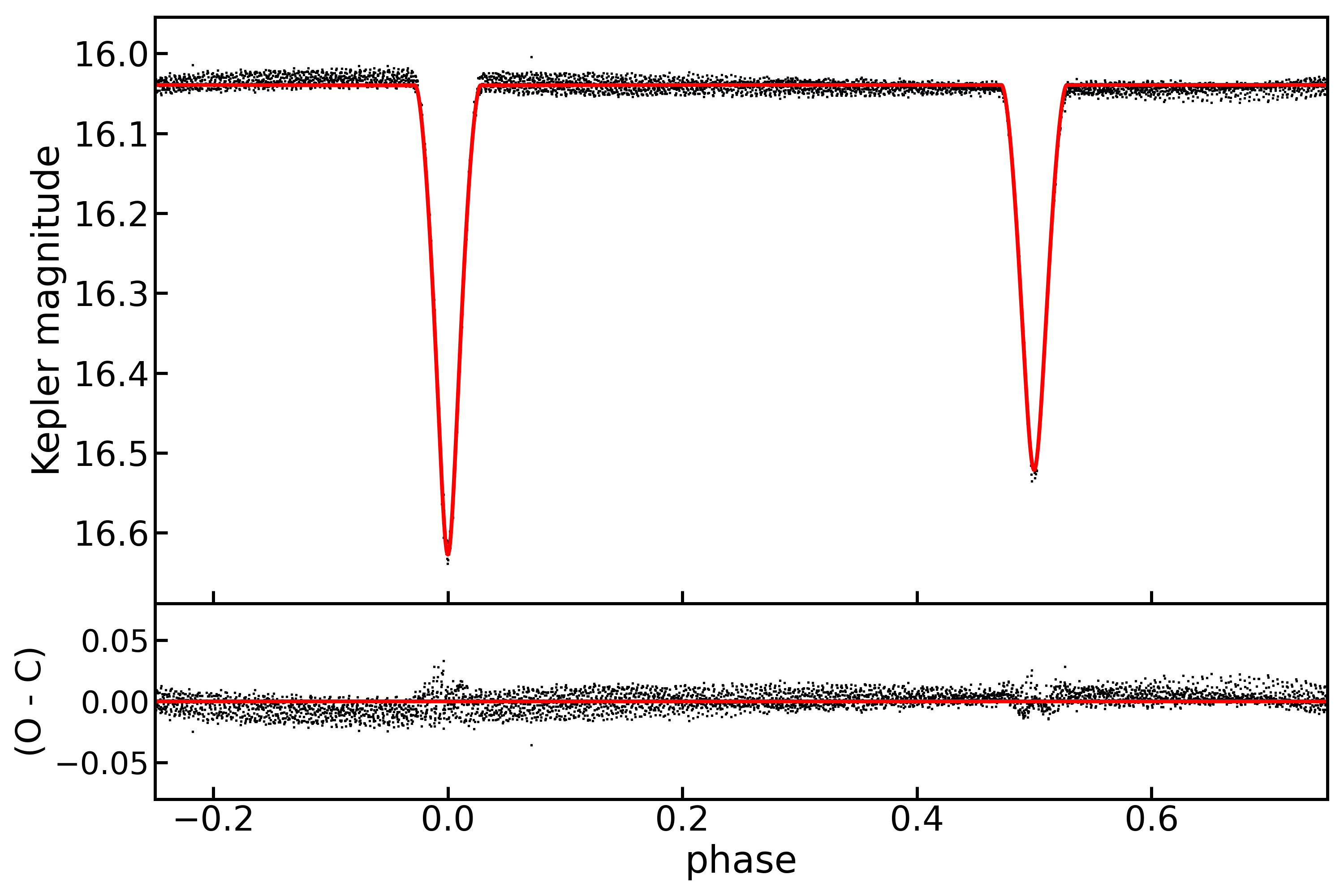}
	\caption{Phase-folded light curve of EB KIC09656543 (tenth quarter). Black dots are the observed data, in Kepler magnitudes. The red solid line shows the best-fitting model from the JKTEBOP code.}
    \label{fig6_phaseLC}
\end{figure}

Convergence was achieved for $94$ DEB systems after four complete iterations. We then performed a Monte Carlo (MC) analysis -- a task included in the JKTEBOP code -- with $10\,000$ iterations per LC, for a robust uncertainties estimation. 
The obtained set of parameters are shown in tables \ref{Tab:param1} and \ref{Tab:param2}, in Appendix \ref{ApenParam}. 
As an example, the phase-folded light curve of KIC09656543 is shown in fig. \ref{fig6_phaseLC}, where we adopted phase zero for the primary eclipse minimum. The best-fitting model is represented by a red solid line.

For these systems, and with the objective of exploring the mass-radius relationship (see Sect. \ref{discuss}), we have estimated their radius values, in units of the solar radius ($R_{\odot}$). The LC-fitting code gives $r_{1}+r_{2}$ and $r_{2}/r_{1}$ -- the sum and the ratio of the stellar radii, respectively -- in units of the binary separation ($a$). To obtain the radii in $R_{\odot}$, we adopted Kepler's Third law to calculate $a$, using the known orbital period and previously derived photometric masses of both components. 

Almost half of these systems presented large error estimates for the photometric masses for the secondary components (with a mean relative error of 62.9\%, approximately), which led to large uncertainties on the calculated radius and a large discrepancy from the expected mass-radius relation for main-sequence stars. 
As mentioned previously, these uncertainties could be related to the the adopted single-epoch photometric measurements, which in turn could be affected by intrinsic stellar variability (see Sect. \ref{KNN}). 
These systems are represented in fig. \ref{fig_HRdiag} by yellow crosses (subsample A). Some of them could have a subgiant component that was not correctly identified by our method based on our ten-colour grid of models. Nevertheless, we believe this is not the general case, as the mentioned EBs are well positioned in the colour-magnitude diagram and have RUWE < 1.4 (for more details, see Sect. \ref{contaminants}). 

Subsample B contains the remaining DEB systems, which in general presented more reliable photometric mass solutions -- the estimated mean relative mass errors for the secondary components are of $\sim$25.5\%. 
They are shown in figure \ref{fig_HRdiag} as red crosses. 
These objects are discussed below, in section \ref{discuss}.

The majority of the non-converged systems presented high variability in the analysed light curve and/or have shallow secondary eclipses. If the LC present intense out-of-eclipse variability, due to stellar spots for instance, they can introduce systematic errors in the LC analysis. In this case, the fitting process may be problematic, especially for low-mass stars with spotted surfaces, since they can result in deeper eclipses or mislead the surface brightness estimation, resulting in wrong $r_{1}+r_{2}$ and $r_{2}/r_{1}$ values \citep[e.g.][]{Irwin2011}. Moreover, the presence of such an intrinsic variability during the eclipse ingress/egress may lead to erroneous radii solutions. 
Under such circumstances, we do not neglect the possibility of cross talk between some of the fitting parameters. Therefore, those systems were rejected on the basis of inconsistent light curve solutions.

\section{Discussion}\label{discuss}

\citet[][]{Prsa2011} -- and the update by \citet[][]{Slawson2011} -- have presented the first EB catalogue from {\it Kepler} database, where they have estimated some orbital parameters like $P_{\rm orb}$, eccentricity and argument of periastron ($e\cos{\omega}$, $e\sin{\omega}$). 
Comparing the analysed sample in this work to their solutions, the majority of them present small eccentricity values, which are consistent in general with our results. They also obtained $r_{1}+r_{2}$ values, the fitting solution for the sum of the stellar radii for the studied systems. The majority of our estimated values are in agreement with those from \citet[][]{Slawson2011}. For comparison, these values are also presented in Appendix \ref{ApenParam} (tables \ref{Tab:param1}, and \ref{Tab:param2}).

\begin{figure}
    \includegraphics[angle=90, width=0.99\columnwidth,trim={2cm 2cm 3cm 3cm},clip]{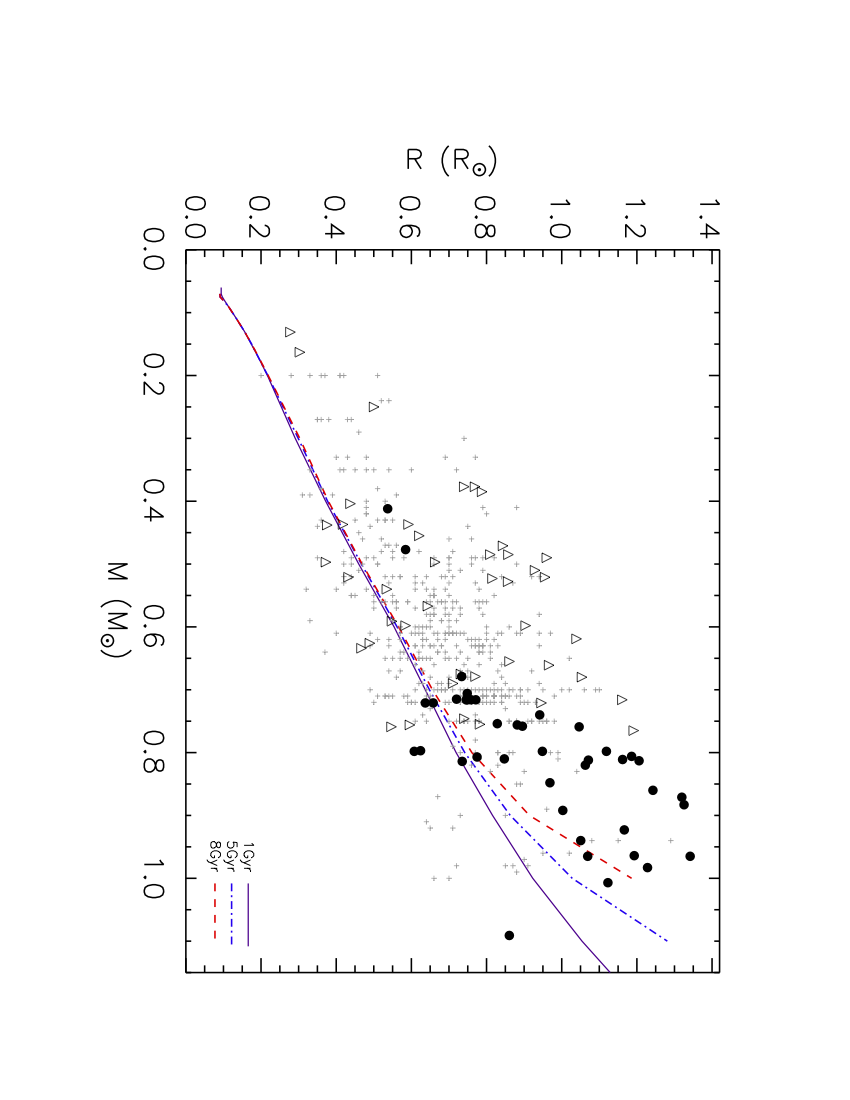}
    \caption{Mass radius diagram for the DEB systems in subsample B. Primary and secondary components are represented by filled circles and open triangles, respectively. \citet{Baraffe2015} standard stellar models ($[M/H]$ = 0, $Y$ = 0.28, $L_{\rm mix} = 1.6 HP$) for $1$, $5$, and $8$ Gyr are represented by solid, dot-dashed, and dashed lines, respectively. DEB candidates from the CSS \citep{Garrido2019} are illustrated as small gray crosses.}
    \label{fig_MRdiag}
\end{figure}

\subsection{The mass-radius diagram}

In order to search for a general trend among the obtained parameters for the DEBs in subsample B, we displayed the radii and masses derived for each binary component in a mass-radius diagram, presented in figure \ref{fig_MRdiag}. The open circles and triangles represent the primary and the secondary components, respectively. 

There seems to be a general trend of radii inflation among our objects, where we derived radii values larger than predicted by stellar evolutionary models, despite the individual uncertainties. 
The standard models from \citet{Baraffe2015} for 1, 5, and 8 Gyr -- computed considering a solar metallicity ($[M/H]$ = 0), with a helium abundance of $Y$ = 0.28, and for a convective mixing length equal to the scale height ($L_{\rm mix} = 1.6 HP$) -- are also illustrated in figure \ref{fig_MRdiag}. For comparison, the results from \citet{Garrido2019} are shown as small gray crosses. We observe that the cloud of points is similarly scattered, as in \citet{Garrido2019}, which may be the result of the larger individual uncertainties obtained by adopting a purely-photometric method (in comparison to smaller errors that can be achieved with spectroscopic methods). It is worth mentioning that other parameters, such as metallicity and age, may also affect the exact locus of individual stars in the mass-radius diagram.

The average difference between the measured stellar radius and expected value for the 5-Gyr model from \citet{Baraffe1998} -- as an estimate of the radius inflation -- appears to be larger for the secondary than for primary components, by over 20\% more inflated. A similar value is estimated from the CSS sample analysed in \citet{Garrido2019} (Fig. \ref{fig_MRdiag}, gray crosses). 

More detailed studies of the final sample presented here, as well as the sample outlined in \citet{Garrido2019} are important to the understanding of the radius inflation. Further search for correlations between the measured inflation and metallicity, age, stellar activity, and/or rotation may hold clues to the mechanism behind the mass-radius anomaly. It may be also important to investigate the possible effects that the adopted sample selection or methodology may have on the obtained results. Interestingly, a recent search for DEBs in CSS database by \citet{Aysses2020} shows no significant inflation among selected targets. We understand the difficulties on deriving stellar physical parameters without spectroscopic data and on developing a model-independent analysis based only on photometry. Therefore, we believe the incongruities found rely on the sample properties, their selection effects and analysis methods.

One may expect a smaller scatter of our DEB systems in the mass-radius diagram when compared to that from CSS, due to the superb {\it Kepler} light curves, which should lead to smaller uncertainties in the estimated stellar radii. However, it is difficult to probe the intrinsic scatter of mass and radii values in both samples. On the other hand, the superb light curves from {\it Kepler} allowed us to resolve other features, such as modulations due to spots for example. This may have an effect on the derived parameters and limit the expected precision, leading to larger uncertainties, as discussed previously.

Our objective was to study homogeneously the whole sample, rejecting unreliable or biased mass or radius estimates, not dealing with each system individually.
For instance, removing the contribution due to spots may improve the radius estimation, however, it may require a case-by-case, time-dependent spot mapping analysis. Further improvements in the light curve analysis will be implemented in order to achieve better radii estimates, i.e. with smaller uncertainties.

\section{Conclusions}\label{concl}

We presented the characterisation of detached eclipsing binary systems from the {\it Kepler} field with low mass components by adopting a purely-photometric method. 

Effective temperatures and photometric masses of individual components we estimated from an extensive multi-colour dataset. We adopted machine learning clustering techniques for the analysis, using a ten-colour binary model grid constructed from the evolutionary stellar models by \citet{Bressan2012}, considering Pan-STARRS and 2MASS broad-band filters. For this new grid, we used a different set of colours (based on the available photometric bands) to generate a more complete set of model binary -- with over $180\,000$ models -- which allowed a better temperature determination than done previously in \citet{Garrido2019}.

For primary components, we obtained temperatures from $3\,152$ to $5\,991$ K, and from $2\,632$ to $5\,991$ K for the secondaries. Differences with published values are greater towards lower temperatures and masses ($T_{\rm eff}<3500$ K; $M_{\star}< 0.4$ M$_\odot$), especially for secondary components, as expected. We intend to explore and add new colours to our grid, covering the infrared region in order to improve the efficiency of our method for binary components in the mentioned low-mass regime.

Previous works also adopted a purely-photometric methodology to derive stellar physical parameters of EB components, using for instance stellar models, available broad-band photometric data and light-curve fitting procedures. \citet{Devor2008} have derived stellar properties from photometric data, although they relied on isochrones for the stellar mass estimation. \citet{Windemuth2019} adopted a similar approach, using Gaia distances as a stellar model constraint. Differently in this work, we adopted semi-empirical values to derive stellar masses. For those systems in common with the sample analysed by \citet{Windemuth2019}, we found a good agreement for the estimated masses, where the median difference is of 20\% for the primary and 45\% for the secondary components, approximately.

We also estimated the radii and orbital parameters from the available {\it Kepler} light curves for $104$ detached EB systems. For that, we used the JKTEBOP code \citep{Southworth2004,Southworth2013} modified by \citet{Coughlin2011} with the asexual genetic algorithm (AGA) by \citet{Canto2009}. 
Spurious values from light-curve modelling were used to reject systems showing high variability -- for instance, due to stellar spots -- and/or eventual subgiant components. Future analysis may account for variability in order to find reliable estimates for such cases.

Despite large individual uncertainties, our results show that there is an inflation trend (of $\approx 20\%$) observed in a mass-radius diagram against theoretical stellar models for low mass regime ($M<1.0 $ M$_\odot$). Nevertheless, additional case-by-case analysis, with spectroscopic masses based on single-target radial velocity measurements, are important to set new boundaries to the problems and further investigate the causes of the radius anomaly in low-mass components of detached EB systems.

\section*{Acknowledgements}

PC acknowledges financial support from the Government of Comunidad Autónoma de Madrid (Spain), via postdoctoral grant ‘Atracción de Talento Investigador’ 2019-T2/TIC-14760. 
MD thanks CNPq funding under grant \#305657. 
JA's contribution to this work is a product of his academic exercise as a professor at the Universidad Militar Nueva Granada, Bogotá, Colombia. 
This research has made use of the Spanish Virtual Observatory (\url{https://svo.cab.inta-csic.es}) project funded by the Spanish Ministry of Science and Innovation/State Agency of Research MCIN/AEI/10.13039/501100011033 through grant PID2020-112949GB-I00 and MDM-2017-0737 at Centro de Astrobiología (CSIC-INTA), Unidad de Excelencia María de Maeztu. 
This research has been partially funded by the Coordena\c c\~ao de Aperfei\c coamento de Pessoal de N\'ivel Superior (CAPES) - Finance Code 001 - Brazil.

This research has made use of the SIMBAD database, operated at CDS, Strasbourg, France. 
This publication makes use of data products from the Two Micron All Sky Survey, which is a joint project of the University of Massachusetts and the Infrared Processing and Analysis Center/California Institute of Technology, funded by the National Aeronautics and Space Administration and the National Science Foundation.

\section*{Data Availability}

The analysed light curves were obtained directly from {\it Kepler} mission archive, available at \url{https://archive.stsci.edu/missions/kepler/lightcurves/}.




\bibliographystyle{mnras.bst} 
\bibliography{Radius_Mass_Kepler_vf.bib}




\appendix

\section{Derived effective temperatures and masses for the 164 identified DEB systems.}\label{ApenTeff}

We present from table \ref{Tab:TeffMass1} to \ref{Tab:TeffMass3}, the derived parameters for the sample of $164$ DEBs identified from the KEBC. The first column shows the {\it Kepler} identification (KIC), followed by the estimated effective temperatures and the masses of each component, which were obtained from a ten-colour grid models, as described in detail in Sect. \ref{charact}.

\begin{table}
\centering
\caption{Derived effective temperatures and masses for the components of the $164$ identified DEB systems from the {\it Kepler} survey. $T_1$ and $M_1$ are the $T_{\rm eff}$ and the stellar mass obtained for the primary component. $T_2$ and $M_2$ are the estimated values for the secondary.}
\begin{small}
\scalebox{0.88}[0.85]{
\begin{tabular}{lcccc}
\hline\hline	

KIC   &	$\mathrm{T}_1$ (K) &   $\mathrm{M}_{1}$ (M$_{\odot}$) &  $\mathrm{T}_2$ (K) &	 $\mathrm{M}_{2}$ (M$_{\odot}$) \\
\hline 

1575690 & 4034 $\pm$ 100 & 0.650 $\pm$ 0.021 & 3208 $\pm$ 509 & 0.236 $\pm$ 0.324\\
2308957 & 5006 $\pm$ 100 & 0.809 $\pm$ 0.023 & 3661 $\pm$ 565 & 0.531 $\pm$ 0.155\\
2447893 & 4972 $\pm$ 100 & 0.802 $\pm$ 0.022 & 3202 $\pm$ 944 & 0.232 $\pm$ 0.441\\
2854948 & 5292 $\pm$ 107 & 0.882 $\pm$ 0.032 & 4694 $\pm$ 355 & 0.752 $\pm$ 0.066\\
2856960 & 4677 $\pm$ 100 & 0.750 $\pm$ 0.016 & 3579 $\pm$ 482 & 0.483 $\pm$ 0.173\\
3122985 & 5364 $\pm$ 100 & 0.910 $\pm$ 0.024 & 3563 $\pm$ 294 & 0.472 $\pm$ 0.127\\
3218683 & 4412 $\pm$ 124 & 0.714 $\pm$ 0.016 & 3416 $\pm$ 283 & 0.371 $\pm$ 0.181\\
3241344 & 5343 $\pm$ 135 & 0.902 $\pm$ 0.037 & 3495 $\pm$ 1112 & 0.426 $\pm$ 0.314\\
3338660 & 5236 $\pm$ 100 & 0.866 $\pm$ 0.029 & 3424 $\pm$ 100 & 0.377 $\pm$ 0.069\\
3344427 & 4429 $\pm$ 100 & 0.716 $\pm$ 0.013 & 3646 $\pm$ 382 & 0.523 $\pm$ 0.125\\
3656322 & 5022 $\pm$ 100 & 0.813 $\pm$ 0.023 & 3546 $\pm$ 124 & 0.461 $\pm$ 0.076\\
3659940 & 5368 $\pm$ 100 & 0.911 $\pm$ 0.024 & 4180 $\pm$ 100 & 0.679 $\pm$ 0.016\\
3662635 & 5328 $\pm$ 109 & 0.892 $\pm$ 0.033 & 4775 $\pm$ 323 & 0.765 $\pm$ 0.065\\
3730067 & 4119 $\pm$ 100 & 0.668 $\pm$ 0.018 & 3181 $\pm$ 486 & 0.220 $\pm$ 0.315\\
3834364 & 5640 $\pm$ 100 & 0.979 $\pm$ 0.046 & 4443 $\pm$ 100 & 0.718 $\pm$ 0.013\\
3848919 & 4952 $\pm$ 100 & 0.798 $\pm$ 0.022 & 4082 $\pm$ 693 & 0.661 $\pm$ 0.104\\
3848972 & 5526 $\pm$ 100 & 0.953 $\pm$ 0.034 & 4501 $\pm$ 376 & 0.725 $\pm$ 0.058\\
3853673 & 5622 $\pm$ 147 & 0.974 $\pm$ 0.060 & 4127 $\pm$ 788 & 0.670 $\pm$ 0.121\\
3973002 & 5580 $\pm$ 113 & 0.965 $\pm$ 0.044 & 3916 $\pm$ 928 & 0.619 $\pm$ 0.158\\
4069063 & 5934 $\pm$ 100 & 1.090 $\pm$ 0.023 & 4530 $\pm$ 102 & 0.729 $\pm$ 0.014\\
4073678 & 5277 $\pm$ 100 & 0.877 $\pm$ 0.029 & 3682 $\pm$ 1384 & 0.543 $\pm$ 0.280\\
4078693 & 5216 $\pm$ 100 & 0.860 $\pm$ 0.028 & 3698 $\pm$ 100 & 0.551 $\pm$ 0.038\\
4174507 & 5425 $\pm$ 100 & 0.929 $\pm$ 0.024 & 3741 $\pm$ 662 & 0.570 $\pm$ 0.143\\
4357272 & 5022 $\pm$ 103 & 0.813 $\pm$ 0.024 & 3419 $\pm$ 887 & 0.373 $\pm$ 0.326\\
4386047 & 5016 $\pm$ 100 & 0.811 $\pm$ 0.023 & 3511 $\pm$ 577 & 0.437 $\pm$ 0.225\\
4480676 & 5052 $\pm$ 100 & 0.819 $\pm$ 0.024 & 3538 $\pm$ 1299 & 0.455 $\pm$ 0.320\\
4484356 & 4951 $\pm$ 100 & 0.798 $\pm$ 0.022 & 3232 $\pm$ 100 & 0.250 $\pm$ 0.064\\
4551328 & 5080 $\pm$ 100 & 0.826 $\pm$ 0.025 & 3496 $\pm$ 1181 & 0.427 $\pm$ 0.323\\
4670267 & 5640 $\pm$ 100 & 0.979 $\pm$ 0.046 & 4612 $\pm$ 100 & 0.740 $\pm$ 0.015\\
4681152 & 5534 $\pm$ 100 & 0.955 $\pm$ 0.035 & 3514 $\pm$ 816 & 0.439 $\pm$ 0.263\\
4757331 & 5021 $\pm$ 100 & 0.812 $\pm$ 0.023 & 3432 $\pm$ 107 & 0.382 $\pm$ 0.074\\
4826439 & 5935 $\pm$ 100 & 1.091 $\pm$ 0.023 & 5765 $\pm$ 100 & 1.017 $\pm$ 0.047\\
4840263 & 5255 $\pm$ 100 & 0.871 $\pm$ 0.029 & 3591 $\pm$ 622 & 0.490 $\pm$ 0.194\\
4902030 & 4196 $\pm$ 100 & 0.682 $\pm$ 0.016 & 2975 $\pm$ 591 & 0.130 $\pm$ 0.344\\
4908495 & 4759 $\pm$ 100 & 0.762 $\pm$ 0.017 & 3294 $\pm$ 618 & 0.288 $\pm$ 0.329\\
5017058 & 5991 $\pm$ 100 & 1.117 $\pm$ 0.010 & 3316 $\pm$ 1325 & 0.303 $\pm$ 0.442\\
5018787 & 5216 $\pm$ 100 & 0.860 $\pm$ 0.028 & 3901 $\pm$ 100 & 0.614 $\pm$ 0.028\\
5022440 & 5515 $\pm$ 145 & 0.951 $\pm$ 0.048 & 3642 $\pm$ 1091 & 0.521 $\pm$ 0.237\\
5036538 & 4431 $\pm$ 100 & 0.716 $\pm$ 0.013 & 3538 $\pm$ 373 & 0.455 $\pm$ 0.162\\
5039441 & 5492 $\pm$ 100 & 0.946 $\pm$ 0.030 & 3625 $\pm$ 131 & 0.511 $\pm$ 0.065\\
5215700 & 5466 $\pm$ 100 & 0.940 $\pm$ 0.027 & 3338 $\pm$ 1111 & 0.317 $\pm$ 0.401\\
5218441 & 5267 $\pm$ 100 & 0.874 $\pm$ 0.029 & 3417 $\pm$ 124 & 0.372 $\pm$ 0.086\\
5636642 & 4947 $\pm$ 100 & 0.797 $\pm$ 0.022 & 4721 $\pm$ 100 & 0.756 $\pm$ 0.016\\
5649956 & 5216 $\pm$ 100 & 0.860 $\pm$ 0.028 & 3557 $\pm$ 100 & 0.468 $\pm$ 0.061\\
5785586 & 5044 $\pm$ 100 & 0.817 $\pm$ 0.024 & 4445 $\pm$ 577 & 0.718 $\pm$ 0.095\\
5802470 & 5498 $\pm$ 100 & 0.947 $\pm$ 0.030 & 3646 $\pm$ 616 & 0.523 $\pm$ 0.169\\
5802486 & 5467 $\pm$ 100 & 0.940 $\pm$ 0.027 & 4185 $\pm$ 228 & 0.680 $\pm$ 0.034\\
6047498 & 5333 $\pm$ 107 & 0.893 $\pm$ 0.033 & 4123 $\pm$ 758 & 0.669 $\pm$ 0.115\\
6058875 & 4429 $\pm$ 100 & 0.716 $\pm$ 0.013 & 3655 $\pm$ 100 & 0.528 $\pm$ 0.047\\
6103049 & 5617 $\pm$ 100 & 0.973 $\pm$ 0.044 & 4333 $\pm$ 343 & 0.703 $\pm$ 0.047\\
6182019 & 5296 $\pm$ 100 & 0.883 $\pm$ 0.030 & 3601 $\pm$ 330 & 0.497 $\pm$ 0.127\\
6187341 & 4665 $\pm$ 100 & 0.748 $\pm$ 0.015 & 3166 $\pm$ 715 & 0.212 $\pm$ 0.396\\
6209347 & 5368 $\pm$ 100 & 0.911 $\pm$ 0.024 & 4158 $\pm$ 100 & 0.675 $\pm$ 0.017\\
6283224 & 5178 $\pm$ 100 & 0.850 $\pm$ 0.027 & 4196 $\pm$ 333 & 0.682 $\pm$ 0.047\\
6311681 & 5084 $\pm$ 100 & 0.827 $\pm$ 0.025 & 3585 $\pm$ 1182 & 0.486 $\pm$ 0.278\\
6452742 & 5443 $\pm$ 100 & 0.934 $\pm$ 0.025 & 4396 $\pm$ 1225 & 0.711 $\pm$ 0.274\\
6469946 & 5040 $\pm$ 100 & 0.817 $\pm$ 0.024 & 4648 $\pm$ 100 & 0.746 $\pm$ 0.015\\
6516874 & 5503 $\pm$ 100 & 0.948 $\pm$ 0.031 & 3612 $\pm$ 164 & 0.503 $\pm$ 0.079\\
6531485 & 5321 $\pm$ 103 & 0.890 $\pm$ 0.031 & 3823 $\pm$ 867 & 0.594 $\pm$ 0.157\\
6543674 & 5509 $\pm$ 100 & 0.949 $\pm$ 0.032 & 3548 $\pm$ 572 & 0.462 $\pm$ 0.206\\
6545018 & 5547 $\pm$ 100 & 0.957 $\pm$ 0.036 & 3642 $\pm$ 1007 & 0.521 $\pm$ 0.225\\
6595662 & 5581 $\pm$ 100 & 0.965 $\pm$ 0.040 & 3641 $\pm$ 100 & 0.521 $\pm$ 0.049\\
6620003 & 3925 $\pm$ 100 & 0.622 $\pm$ 0.026 & 3174 $\pm$ 399 & 0.217 $\pm$ 0.262\\
6665064 & 5640 $\pm$ 100 & 0.979 $\pm$ 0.046 & 3971 $\pm$ 100 & 0.634 $\pm$ 0.024\\
6695889 & 5801 $\pm$ 100 & 1.031 $\pm$ 0.045 & 3607 $\pm$ 158 & 0.500 $\pm$ 0.079\\
6697716 & 4735 $\pm$ 100 & 0.759 $\pm$ 0.017 & 4735 $\pm$ 100 & 0.759 $\pm$ 0.017\\
6706287 & 4998 $\pm$ 100 & 0.807 $\pm$ 0.023 & 3624 $\pm$ 100 & 0.510 $\pm$ 0.052\\
6778050 & 5054 $\pm$ 100 & 0.820 $\pm$ 0.024 & 4178 $\pm$ 100 & 0.679 $\pm$ 0.016\\
6862603 & 5654 $\pm$ 103 & 0.983 $\pm$ 0.047 & 3892 $\pm$ 819 & 0.611 $\pm$ 0.144\\
6863840 & 5030 $\pm$ 100 & 0.814 $\pm$ 0.024 & 3851 $\pm$ 494 & 0.598 $\pm$ 0.106\\

 \hline\hline
\end{tabular}}
\label{Tab:TeffMass1}
\end{small}
\end{table}

\begin{table}
\centering
\caption{Continued from Table \ref{Tab:TeffMass1}.}
\begin{small}
\scalebox{0.88}[0.85]{
\begin{tabular}{lcccc}
\hline\hline	

KIC   &	$\mathrm{T}_1$ (K) &   $\mathrm{M}_{1}$ (M$_{\odot}$) &  $\mathrm{T}_2$ (K) &	 $\mathrm{M}_{2}$ (M$_{\odot}$) \\
\hline 

7025540 & 4429 $\pm$ 100 & 0.716 $\pm$ 0.013 & 3602 $\pm$ 377 & 0.497 $\pm$ 0.139\\
7174617 & 4957 $\pm$ 100 & 0.799 $\pm$ 0.022 & 3577 $\pm$ 100 & 0.481 $\pm$ 0.059\\
7212066 & 5262 $\pm$ 100 & 0.873 $\pm$ 0.029 & 3682 $\pm$ 141 & 0.543 $\pm$ 0.052\\
7220320 & 5251 $\pm$ 100 & 0.870 $\pm$ 0.029 & 3584 $\pm$ 100 & 0.486 $\pm$ 0.058\\
7295570 & 5169 $\pm$ 100 & 0.848 $\pm$ 0.027 & 3337 $\pm$ 100 & 0.317 $\pm$ 0.069\\
7374746 & 5019 $\pm$ 100 & 0.812 $\pm$ 0.023 & 4714 $\pm$ 100 & 0.755 $\pm$ 0.016\\
7377033 & 4984 $\pm$ 100 & 0.804 $\pm$ 0.022 & 4747 $\pm$ 100 & 0.761 $\pm$ 0.017\\
7552344 & 5012 $\pm$ 100 & 0.810 $\pm$ 0.023 & 3290 $\pm$ 818 & 0.286 $\pm$ 0.380\\
7605600 & 3698 $\pm$ 100 & 0.551 $\pm$ 0.038 & 3060 $\pm$ 401 & 0.162 $\pm$ 0.241\\
7671594 & 3344 $\pm$ 159 & 0.321 $\pm$ 0.111 & 2893 $\pm$ 335 & 0.108 $\pm$ 0.139\\
7769072 & 4653 $\pm$ 100 & 0.746 $\pm$ 0.015 & 3648 $\pm$ 486 & 0.524 $\pm$ 0.147\\
7798259 & 4721 $\pm$ 100 & 0.756 $\pm$ 0.016 & 3941 $\pm$ 100 & 0.626 $\pm$ 0.026\\
7830321 & 5169 $\pm$ 100 & 0.848 $\pm$ 0.027 & 3518 $\pm$ 100 & 0.442 $\pm$ 0.065\\
7842610 & 5142 $\pm$ 133 & 0.841 $\pm$ 0.036 & 3659 $\pm$ 1123 & 0.531 $\pm$ 0.236\\
7947631 & 4841 $\pm$ 100 & 0.776 $\pm$ 0.019 & 3577 $\pm$ 519 & 0.481 $\pm$ 0.182\\
8094140 & 4177 $\pm$ 100 & 0.679 $\pm$ 0.016 & 3463 $\pm$ 285 & 0.404 $\pm$ 0.169\\
8097825 & 5262 $\pm$ 100 & 0.873 $\pm$ 0.029 & 3640 $\pm$ 100 & 0.520 $\pm$ 0.050\\
8111381 & 5473 $\pm$ 110 & 0.941 $\pm$ 0.031 & 4623 $\pm$ 891 & 0.742 $\pm$ 0.208\\
8127648 & 4353 $\pm$ 129 & 0.706 $\pm$ 0.017 & 3511 $\pm$ 539 & 0.437 $\pm$ 0.216\\
8145789 & 4731 $\pm$ 100 & 0.758 $\pm$ 0.017 & 4058 $\pm$ 1143 & 0.655 $\pm$ 0.201\\
8211824 & 4612 $\pm$ 100 & 0.740 $\pm$ 0.015 & 2683 $\pm$ 227 & 0.081 $\pm$ 0.031\\
8231877 & 5021 $\pm$ 100 & 0.812 $\pm$ 0.023 & 3598 $\pm$ 153 & 0.494 $\pm$ 0.079\\
8244173 & 4425 $\pm$ 100 & 0.715 $\pm$ 0.013 & 2972 $\pm$ 732 & 0.129 $\pm$ 0.424\\
8279765 & 5578 $\pm$ 100 & 0.964 $\pm$ 0.040 & 4471 $\pm$ 100 & 0.721 $\pm$ 0.013\\
8288719 & 4985 $\pm$ 100 & 0.804 $\pm$ 0.023 & 3559 $\pm$ 113 & 0.469 $\pm$ 0.068\\
8397675 & 5381 $\pm$ 100 & 0.916 $\pm$ 0.023 & 4430 $\pm$ 100 & 0.716 $\pm$ 0.013\\
8411947 & 4974 $\pm$ 100 & 0.802 $\pm$ 0.022 & 3218 $\pm$ 990 & 0.241 $\pm$ 0.443\\
8435247 & 4798 $\pm$ 105 & 0.769 $\pm$ 0.019 & 3472 $\pm$ 716 & 0.410 $\pm$ 0.270\\
8444552 & 5310 $\pm$ 100 & 0.887 $\pm$ 0.030 & 3625 $\pm$ 100 & 0.511 $\pm$ 0.052\\
8616873 & 5287 $\pm$ 100 & 0.880 $\pm$ 0.030 & 3916 $\pm$ 750 & 0.619 $\pm$ 0.129\\
8620561 & 4431 $\pm$ 100 & 0.716 $\pm$ 0.013 & 3031 $\pm$ 575 & 0.150 $\pm$ 0.350\\
8879915 & 5416 $\pm$ 100 & 0.927 $\pm$ 0.024 & 3396 $\pm$ 1751 & 0.357 $\pm$ 0.485\\
8949316 & 3570 $\pm$ 100 & 0.477 $\pm$ 0.060 & 3063 $\pm$ 322 & 0.163 $\pm$ 0.187\\
8971432 & 4962 $\pm$ 100 & 0.800 $\pm$ 0.022 & 3493 $\pm$ 558 & 0.425 $\pm$ 0.229\\
9005854 & 5467 $\pm$ 100 & 0.940 $\pm$ 0.027 & 4185 $\pm$ 228 & 0.680 $\pm$ 0.034\\
9053086 & 5691 $\pm$ 100 & 0.993 $\pm$ 0.048 & 3722 $\pm$ 254 & 0.562 $\pm$ 0.074\\
9110346 & 4506 $\pm$ 100 & 0.726 $\pm$ 0.014 & 3104 $\pm$ 778 & 0.181 $\pm$ 0.427\\
9288786 & 4178 $\pm$ 100 & 0.679 $\pm$ 0.016 & 2875 $\pm$ 518 & 0.104 $\pm$ 0.250\\
9346655 & 4177 $\pm$ 100 & 0.679 $\pm$ 0.016 & 3463 $\pm$ 285 & 0.404 $\pm$ 0.169\\
9411943 & 5525 $\pm$ 100 & 0.953 $\pm$ 0.034 & 3693 $\pm$ 194 & 0.549 $\pm$ 0.061\\
9474485 & 4425 $\pm$ 100 & 0.715 $\pm$ 0.013 & 3424 $\pm$ 567 & 0.377 $\pm$ 0.263\\
9540450 & 5402 $\pm$ 100 & 0.923 $\pm$ 0.023 & 3436 $\pm$ 150 & 0.385 $\pm$ 0.102\\
9593759 & 4534 $\pm$ 100 & 0.730 $\pm$ 0.014 & 3545 $\pm$ 411 & 0.460 $\pm$ 0.170\\
9596187 & 5658 $\pm$ 108 & 0.984 $\pm$ 0.050 & 3078 $\pm$ 458 & 0.169 $\pm$ 0.285\\
9597095 & 5935 $\pm$ 100 & 1.091 $\pm$ 0.023 & 5798 $\pm$ 100 & 1.030 $\pm$ 0.045\\
9639491 & 4721 $\pm$ 100 & 0.756 $\pm$ 0.016 & 3430 $\pm$ 100 & 0.381 $\pm$ 0.069\\
9641018 & 5593 $\pm$ 100 & 0.968 $\pm$ 0.041 & 4601 $\pm$ 673 & 0.739 $\pm$ 0.137\\
9656543 & 5012 $\pm$ 100 & 0.810 $\pm$ 0.023 & 4156 $\pm$ 789 & 0.675 $\pm$ 0.121\\
9664215 & 5631 $\pm$ 100 & 0.977 $\pm$ 0.045 & 4302 $\pm$ 324 & 0.698 $\pm$ 0.044\\
9665086 & 5090 $\pm$ 121 & 0.828 $\pm$ 0.031 & 4037 $\pm$ 636 & 0.651 $\pm$ 0.098\\
9761199 & 4014 $\pm$ 100 & 0.645 $\pm$ 0.022 & 3197 $\pm$ 511 & 0.230 $\pm$ 0.326\\
9784230 & 4471 $\pm$ 100 & 0.721 $\pm$ 0.013 & 3583 $\pm$ 311 & 0.485 $\pm$ 0.127\\
9813678 & 4995 $\pm$ 100 & 0.807 $\pm$ 0.023 & 3377 $\pm$ 272 & 0.344 $\pm$ 0.181\\
9834719 & 5879 $\pm$ 100 & 1.065 $\pm$ 0.034 & 3733 $\pm$ 975 & 0.567 $\pm$ 0.188\\
9912977 & 5169 $\pm$ 100 & 0.848 $\pm$ 0.027 & 3741 $\pm$ 173 & 0.570 $\pm$ 0.048\\
9944201 & 4747 $\pm$ 100 & 0.761 $\pm$ 0.017 & 3571 $\pm$ 100 & 0.477 $\pm$ 0.060\\
9945280 & 5471 $\pm$ 100 & 0.941 $\pm$ 0.028 & 3370 $\pm$ 171 & 0.339 $\pm$ 0.118\\
10014830 & 5038 $\pm$ 123 & 0.816 $\pm$ 0.030 & 3675 $\pm$ 1016 & 0.539 $\pm$ 0.213\\
10068030 & 5581 $\pm$ 100 & 0.965 $\pm$ 0.040 & 4430 $\pm$ 100 & 0.716 $\pm$ 0.013\\
10083623 & 5381 $\pm$ 100 & 0.916 $\pm$ 0.023 & 3941 $\pm$ 100 & 0.626 $\pm$ 0.026\\
10090246 & 5425 $\pm$ 100 & 0.930 $\pm$ 0.024 & 3617 $\pm$ 208 & 0.506 $\pm$ 0.089\\
10129482 & 4702 $\pm$ 100 & 0.754 $\pm$ 0.016 & 3734 $\pm$ 352 & 0.567 $\pm$ 0.094\\
10189523 & 5127 $\pm$ 133 & 0.837 $\pm$ 0.035 & 3950 $\pm$ 735 & 0.629 $\pm$ 0.122\\
10257903 & 5565 $\pm$ 100 & 0.961 $\pm$ 0.038 & 4387 $\pm$ 766 & 0.710 $\pm$ 0.134\\
10264202 & 5169 $\pm$ 100 & 0.848 $\pm$ 0.027 & 3562 $\pm$ 100 & 0.471 $\pm$ 0.061\\
10346522 & 5292 $\pm$ 100 & 0.882 $\pm$ 0.030 & 3868 $\pm$ 360 & 0.599 $\pm$ 0.088\\
10363300 & 4990 $\pm$ 100 & 0.806 $\pm$ 0.023 & 3513 $\pm$ 115 & 0.438 $\pm$ 0.074\\
10468514 & 4962 $\pm$ 100 & 0.800 $\pm$ 0.022 & 3091 $\pm$ 824 & 0.175 $\pm$ 0.443\\
10556578 & 4952 $\pm$ 100 & 0.798 $\pm$ 0.022 & 3489 $\pm$ 100 & 0.422 $\pm$ 0.067\\
10666230 & 5831 $\pm$ 100 & 1.043 $\pm$ 0.041 & 4426 $\pm$ 411 & 0.715 $\pm$ 0.060\\
10728219 & 5935 $\pm$ 100 & 1.091 $\pm$ 0.023 & 4430 $\pm$ 100 & 0.716 $\pm$ 0.013\\
10794405 & 5490 $\pm$ 106 & 0.945 $\pm$ 0.032 & 4949 $\pm$ 255 & 0.797 $\pm$ 0.060\\
11076176 & 4981 $\pm$ 100 & 0.804 $\pm$ 0.022 & 3840 $\pm$ 700 & 0.597 $\pm$ 0.134\\

\hline\hline
\end{tabular}}
\label{Tab:TeffMass2}
\end{small}
\end{table}

\begin{table}
\centering
\caption{Continued from Table \ref{Tab:TeffMass2}.}
\begin{small}
\scalebox{0.88}[0.85]{
\begin{tabular}{lcccc}
\hline\hline	

KIC   &	$\mathrm{T}_1$ (K) &   $\mathrm{M}_{1}$ (M$_{\odot}$) &  $\mathrm{T}_2$ (K) &	 $\mathrm{M}_{2}$ (M$_{\odot}$) \\
\hline 

11134079 & 5346 $\pm$ 110 & 0.903 $\pm$ 0.028 & 4483 $\pm$ 319 & 0.723 $\pm$ 0.047\\
11147276 & 5657 $\pm$ 100 & 0.983 $\pm$ 0.047 & 3676 $\pm$ 830 & 0.540 $\pm$ 0.186\\
11198068 & 5026 $\pm$ 100 & 0.813 $\pm$ 0.024 & 4648 $\pm$ 100 & 0.746 $\pm$ 0.015\\
11200773 & 5991 $\pm$ 100 & 1.117 $\pm$ 0.010 & 4747 $\pm$ 100 & 0.761 $\pm$ 0.017\\
11228612 & 5734 $\pm$ 110 & 1.007 $\pm$ 0.052 & 4248 $\pm$ 722 & 0.690 $\pm$ 0.111\\
11304987 & 5639 $\pm$ 100 & 0.979 $\pm$ 0.046 & 4172 $\pm$ 100 & 0.678 $\pm$ 0.016\\
11404698 & 4178 $\pm$ 100 & 0.679 $\pm$ 0.016 & 2875 $\pm$ 518 & 0.104 $\pm$ 0.250\\
11455795 & 4440 $\pm$ 100 & 0.717 $\pm$ 0.013 & 4440 $\pm$ 100 & 0.717 $\pm$ 0.013\\
11457191 & 5640 $\pm$ 100 & 0.979 $\pm$ 0.046 & 3971 $\pm$ 100 & 0.634 $\pm$ 0.024\\
11564013 & 5434 $\pm$ 100 & 0.932 $\pm$ 0.025 & 3336 $\pm$ 829 & 0.316 $\pm$ 0.361\\
11616200 & 5513 $\pm$ 250 & 0.950 $\pm$ 0.082 & 3851 $\pm$ 888 & 0.598 $\pm$ 0.161\\
12004834 & 3475 $\pm$ 100 & 0.412 $\pm$ 0.068 & 2976 $\pm$ 376 & 0.131 $\pm$ 0.196\\
12010534 & 5713 $\pm$ 100 & 1.000 $\pm$ 0.049 & 4325 $\pm$ 207 & 0.702 $\pm$ 0.028\\
12022517 & 5216 $\pm$ 100 & 0.860 $\pm$ 0.028 & 3424 $\pm$ 100 & 0.377 $\pm$ 0.069\\
12023089 & 4955 $\pm$ 116 & 0.798 $\pm$ 0.026 & 3809 $\pm$ 815 & 0.591 $\pm$ 0.151\\
12109575 & 4471 $\pm$ 100 & 0.721 $\pm$ 0.013 & 3696 $\pm$ 158 & 0.550 $\pm$ 0.049\\
12109845 & 4612 $\pm$ 100 & 0.740 $\pm$ 0.015 & 3971 $\pm$ 100 & 0.634 $\pm$ 0.024\\
12365000 & 5025 $\pm$ 100 & 0.813 $\pm$ 0.024 & 3688 $\pm$ 215 & 0.546 $\pm$ 0.069\\
12418662 & 5368 $\pm$ 100 & 0.911 $\pm$ 0.024 & 4135 $\pm$ 213 & 0.671 $\pm$ 0.034\\
12418816 & 4471 $\pm$ 100 & 0.721 $\pm$ 0.013 & 3583 $\pm$ 311 & 0.485 $\pm$ 0.127\\
12553806 & 5962 $\pm$ 100 & 1.104 $\pm$ 0.017 & 3314 $\pm$ 1319 & 0.302 $\pm$ 0.442\\

\hline\hline
\end{tabular}}
\label{Tab:TeffMass3}
\end{small}
\end{table}

\section{Derived parameters from light curve fitting for 94 DEB systems.}\label{ApenParam}

We present in tables \ref{Tab:param1} and \ref{Tab:param2} the derived parameters from the light curve fitting with JKTEBOP+AGA, for a list of $94$ DEBs from the KEBC, as described in detail in Sect. \ref{model}. 

The {\it Kepler} identification (KIC) is presented in column 1. Columns 2 and 3 show the sum and the ratio of stellar radii, $r1+r2$ and $r2/r1$, respectively. The central surface brightness ratio, $J=J_2/J_1$, is shown in column 4. The estimated orbital inclination and eccentricity are presented in columns 5 and 6, respectively. Column 7 shows the orbital period ($P_{\rm orb}$) from \citet{Kirk2016}, which was kept fixed during the fitting procedure. Column 8 shows the reference time (epoch) of primary minimum ($T_{\rm 0}$). The radii values, $R_{1}$ and $R_{2}$, presented in columns 9 and 10 were calculated by using Kepler's third law. Column 11 shows the sum of the radii, $(r1+r2)_{pub}$, obtained by \citet{Slawson2011}, for comparison. Column 12 presents the subsample, A or B (see Sect.\ref{discuss} for details).

\begin{table*}
\centering
\caption{Orbital and Physical parameters obtained from LC fitting with JKTEBOP for $94$ DEBs, with orbital period shorter $4$ days, identified from the KEBC. The radii values presented in columns 9 and 10 were estimated using Kepler's third law. Column 11 shows the sum of the radii obtained by \citet{Slawson2011}.}
\begin{tabular}{lccccccccccc}
\hline\hline	

KIC & $r1+r2$  & $r2/r1$  & $J$ & $i$ & $ecc$ & $P_{\rm orb}$ & $T_{0}$ & $R_{1}$ & $R_{2}$ & $(r1+r2)_{pub}$ & subsample \\
  
  & ($a$) &  &  & ($^{\circ}$) &  & (days) & (MJD-2400000) & (R$_{\odot}$) &(R$_{\odot}$) & ($a$) & \\
\hline

  2308957 & 0.4725 & 1.1628 & 0.8652 & 80.94 & 0.0 & 2.2196838 & 54965.16547 & 1.723 & 2.004 & 0.448& A \\
  2447893 & 0.5021 & 0.7926 & 0.7119 & 72.32 & 0.002 & 0.66162 & 54965.1142 & 0.904 & 0.717 & 0.56& A \\
  2854948 & 0.2154 & 0.1653 & 0.0 & 89.84 & 0.132 & 0.9743044 & 55002.41043 & 0.9 & 0.149 & - & A \\
  3218683 & 0.6641 & 0.5164 & 0.6853 & 76.52 & 0.002 & 0.7716695 & 54965.11791 & 1.592 & 0.822 & 0.684& A \\
  3338660 & 0.6328 & 0.8408 & 0.0942 & 79.35 & 0.004 & 1.8733805 & 55002.26338 & 2.362 & 1.986 & - & A \\
  3344427 & 0.4675 & 1.0575 & 0.5449 & 78.78 & 0.0 & 0.6517851 & 55000.00332 & 0.771 & 0.816 & - & B \\
  3659940 & 0.2555 & 0.2137 & 0.6779 & 82.19 & 0.011 & 0.8962698 & 54965.56925 & 0.96 & 0.205 & 0.391& A \\
  3662635 & 0.4598 & 1.1882 & 1.0377 & 70.79 & 0.0 & 0.9393755 & 54965.01636 & 1.003 & 1.192 & 0.532& B \\
  3730067 & 0.6244 & 1.1254 & 0.5661 & 72.93 & 0.007 & 0.2940786 & 54964.88723 & 0.525 & 0.591 & 0.668& A \\
  3848919 & 0.424 & 0.8638 & 0.9086 & 84.43 & 0.0 & 1.0472603 & 54964.76661 & 1.119 & 0.967 & 0.412& B \\
  3973002 & 0.1933 & 0.7748 & 0.8305 & 81.82 & 0.0 & 3.9841974 & 54967.88329 & 1.342 & 1.04 & 0.191& B \\
  4386047 & 0.1716 & 0.3599 & 0.0753 & 87.62 & 0.002 & 2.90069 & 55001.19665 & 1.162 & 0.418 & - & B \\
  4484356 & 0.3102 & 0.528 & 0.8263 & 79.62 & 0.0 & 1.1441593 & 54965.01868 & 0.949 & 0.501 & 0.572& B \\
  4670267 & 0.4928 & 0.6643 & 0.8283 & 77.68 & 0.0 & 2.0060984 & 54966.37549 & 2.372 & 1.576 & 0.501& A \\
  4681152 & 0.7014 & 7.7565 & 0.5339 & 9.82 & 0.95 & 1.8359219 & 54954.06658 & 0.564 & 4.376 & 0.239& A \\
  4826439 & 0.3719 & 1.004 & 1.0104 & 87.6 & 0.0 & 2.4742946 & 54965.65825 & 1.828 & 1.835 & 0.372& B \\
  4840263 & 0.3172 & 0.7278 & 0.4805 & 78.26 & 0.0 & 1.9156478 & 54966.37262 & 1.32 & 0.96 & 0.316& B \\
  4902030 & 0.2165 & 0.8547 & 0.6284 & 88.43 & 0.0 & 1.757606 & 54965.26791 & 0.667 & 0.57 & 0.21& A \\
  5017058 & 0.3084 & 0.802 & 0.9182 & 79.04 & 0.0 & 2.3238948 & 54955.91658 & 1.417 & 1.136 & 0.301& A \\
  5036538 & 0.1872 & 0.8296 & 0.8992 & 88.81 & 0.0 & 2.1220164 & 54965.52097 & 0.749 & 0.621 & 0.186& B \\
  5215700 & 0.4533 & 0.8021 & 0.9239 & 75.58 & 0.001 & 1.3124021 & 55186.23876 & 1.368 & 1.098 & - & A \\
  5636642 & 0.2624 & 0.9547 & 1.1119 & 76.46 & 0.0 & 0.9335 & 54999.62627 & 0.624 & 0.596 & - & B \\
  5649956 & 0.29 & 0.6924 & 0.74 & 76.99 & 0.001 & 2.41574 & 55003.95632 & 1.426 & 0.987 & - & A \\
  5785586 & 0.2098 & 0.156 & 2.0E-4 & 89.99 & 0.006 & 0.4596214 & 54964.81669 & 0.524 & 0.082 & 0.377& A \\
  6058875 & 0.3268 & 1.1499 & 0.806 & 81.98 & 0.0 & 1.1298682 & 55002.0728 & 0.746 & 0.858 & - & B \\
  6103049 & 0.2422 & 0.1629 & 0.6026 & 89.99 & 0.01 & 0.6431712 & 54964.88903 & 0.775 & 0.126 & 0.425& A \\
  6182019 & 0.1526 & 0.2812 & 0.0539 & 86.23 & 0.0 & 3.6649654 & 55003.80231 & 1.326 & 0.373 & 0.206& B \\
  6209347 & 0.1937 & 0.265 & 0.4824 & 82.88 & 0.0 & 2.1365771 & 55004.48388 & 1.246 & 0.33 & 0.211& A \\
  6516874 & 0.4984 & 0.8819 & 0.9794 & 68.63 & 0.006 & 0.9163269 & 55001.92442 & 1.189 & 1.049 & - & A \\
  6531485 & 0.1717 & 0.1159 & 0.7726 & 89.99 & 0.0 & 0.6769904 & 54964.80143 & 0.569 & 0.066 & 0.228& A \\
  6543674 & 0.3318 & 0.7663 & 1.0311 & 87.67 & 0.0 & 2.3910305 & 54965.30428 & 1.584 & 1.214 & 0.313& A \\
  6545018 & 0.2175 & 0.574 & 0.8427 & 86.93 & 0.002 & 3.9914603 & 54965.83609 & 1.665 & 0.956 & 0.223& B \\
  6595662 & 0.1622 & 0.4048 & 0.108 & 87.42 & 0.0 & 2.6805144 & 54965.59817 & 1.069 & 0.433 & 0.204& B \\
  6620003 & 0.1589 & 0.9006 & 1.1729 & 82.22 & 0.0 & 3.4285506 & 54966.31624 & 0.754 & 0.679 & 0.143& A \\
  6695889 & 0.3754 & 1.2786 & 0.2674 & 76.71 & 0.001 & 1.1065608 & 54965.37458 & 0.854 & 1.092 & 0.385& A \\
  6697716 & 0.2582 & 0.523 & 0.5323 & 83.27 & 0.001 & 1.443249 & 54965.58148 & 1.046 & 0.547 & 0.267& B \\
  6706287 & 0.1988 & 1.1991 & 0.6927 & 86.59 & 0.0 & 2.5353877 & 54966.38831 & 0.775 & 0.929 & 0.196& B \\
  6778050 & 0.3954 & 0.7252 & 0.9327 & 81.69 & 0.0 & 0.945828 & 54965.56604 & 1.063 & 0.771 & 0.392& B \\
  6863840 & 0.1414 & 1.2296 & 0.768 & 88.81 & 0.0 & 3.8527339 & 54964.7474 & 0.735 & 0.904 & 0.132& B \\
  7025540 & 0.1905 & 0.8745 & 0.7868 & 88.35 & 0.001 & 2.1482069 & 55279.40867 & 0.759 & 0.664 & - & B \\
  7212066 & 0.1176 & 0.1978 & 0.18 & 85.58 & 0.002 & 3.8404858 & 55186.35611 & 1.137 & 0.225 & - & A \\
  7374746 & 0.1941 & 0.7307 & 0.7725 & 84.03 & 0.0 & 2.7338924 & 55186.65008 & 1.071 & 0.783 & - & B \\
  7605600 & 0.1234 & 1.3857 & 0.3307 & 85.25 & 0.0 & 3.3261926 & 55006.24447 & 0.433 & 0.6 & 0.129& A \\
  7798259 & 0.2028 & 0.5552 & 0.3856 & 86.01 & 0.001 & 1.7342219 & 54965.83367 & 0.882 & 0.489 & 0.117& B \\
  7947631 & 0.0806 & 0.1051 & 0.9778 & 88.03 & 0.0 & 2.5165557 & 54966.96154 & 0.612 & 0.064 & 0.152& A \\
  8094140 & 0.3421 & 0.5969 & 0.3532 & 85.09 & 0.002 & 0.7064291 & 54965.14719 & 0.734 & 0.438 & 0.249& B \\
  8097825 & 0.3012 & 0.4278 & 0.7483 & 82.82 & 0.001 & 2.9368523 & 54954.88717 & 2.032 & 0.869 & 0.325& A \\
  8127648 & 0.1892 & 0.7914 & 0.6727 & 83.32 & 0.0 & 2.0469445 & 54999.49888 & 0.749 & 0.592 & - & B \\
  8145789 & 0.2644 & 0.9617 & 0.7824 & 76.5 & 0.0 & 1.6706295 & 54964.95712 & 0.895 & 0.861 & 0.217& B \\
  8211824 & 0.6328 & 0.1197 & 0.0 & 89.79 & 0.207 & 0.8411262 & 55000.11865 & 1.983 & 0.237 & - & A \\
  8231877 & 0.0846 & 0.1167 & 0.6226 & 89.07 & 0.0 & 2.6155284 & 54964.77882 & 0.661 & 0.077 & 0.161& A \\
  8244173 & 0.1905 & 1.1879 & 0.978 & 88.15 & 0.0 & 2.1841323 & 55185.70006 & 0.582 & 0.692 & - & A \\
  8279765 & 0.2174 & 0.7933 & 0.296 & 79.23 & 0.001 & 2.7577936 & 54965.47576 & 1.193 & 0.946 & 0.179& B \\
  8411947 & 0.2705 & 1.0694 & 0.5972 & 86.62 & 0.001 & 1.7976748 & 54966.03359 & 0.824 & 0.881 & 0.216& A \\
  8444552 & 0.2346 & 0.2448 & 0.8686 & 82.61 & 0.001 & 1.1780996 & 54964.59007 & 0.988 & 0.242 & 0.365& A \\
  8620561 & 0.375 & 0.8638 & 0.4579 & 80.2 & 0.0 & 0.7820467 & 55002.21229 & 0.685 & 0.591 & - & A \\
  8879915 & 0.1607 & 0.835 & 0.5404 & 89.24 & 0.0 & 3.442627 & 54965.53234 & 0.913 & 0.762 & 0.159& A \\
  8949316 & 0.3428 & 0.519 & 0.7125 & 73.99 & 0.0 & 0.6043566 & 54964.68768 & 0.585 & 0.303 & - & B \\
  8971432 & 0.2293 & 0.1514 & 0.0547 & 89.99 & 0.008 & 0.6243873 & 54964.79601 & 0.655 & 0.099 & 0.303& A \\

\hline\hline
 
\end{tabular}
\label{Tab:param1}
\end{table*}

\begin{table*}
\centering
\caption{Continued from Table \ref{Tab:param1}.}
\begin{tabular}{lccccccccccc}
\hline\hline	

KIC & $r1+r2$  & $r2/r1$  & $J$ & $i$ & $ecc$ & $P_{\rm orb}$ & $T_{0}$ & $R_{1}$ & $R_{2}$ & $(r1+r2)_{pub}$ & subsample \\
  
  & ($a$) &  &  & ($^{\circ}$) &  & (days) & (MJD-2400000) & (R$_{\odot}$) &(R$_{\odot}$) & ($a$) & \\
\hline

  9005854 & 0.1756 & 1.0032 & 0.3666 & 83.26 & 0.0 & 3.7804542 & 54968.45923 & 1.051 & 1.054 & 0.178& B \\
  9053086 & 0.3074 & 0.2054 & 0.2372 & 78.8 & 0.002 & 1.2748413 & 55000.31839 & 1.461 & 0.3 & - & A \\
  9110346 & 0.2401 & 0.9123 & 0.8403 & 86.04 & 0.0 & 1.7905531 & 55002.22285 & 0.754 & 0.688 & - & A \\
  9474485 & 0.3318 & 1.0285 & 0.853 & 85.55 & 0.0 & 1.0251632 & 54965.29343 & 0.72 & 0.741 & 0.33& B \\
  9540450 & 0.2548 & 0.6757 & 0.5005 & 85.37 & 0.0 & 2.1547127 & 54966.75556 & 1.167 & 0.788 & 0.25& B \\
  9597095 & 0.1267 & 0.1905 & 0.1256 & 87.32 & 0.0 & 2.7456399 & 54967.45334 & 1.127 & 0.215 & 0.22& A \\
  9639491 & 0.7386 & 1.1519 & 0.9495 & 65.54 & 0.0 & 0.3445456 & 54964.92639 & 0.74 & 0.853 & - & A \\
  9656543 & 0.1766 & 0.8637 & 0.8937 & 88.06 & 0.0 & 2.544565 & 55002.40155 & 0.847 & 0.732 & - & B \\
  9784230 & 0.3812 & 1.2311 & 0.9548 & 81.34 & 0.0 & 0.7981975 & 55000.48804 & 0.658 & 0.81 & - & B \\
  9813678 & 0.2148 & 0.1337 & 0.6809 & 89.99 & 0.001 & 0.5050786 & 55000.28706 & 0.53 & 0.071 & - & A \\
  9912977 & 0.4766 & 0.8887 & 0.9461 & 79.33 & 0.0 & 1.8878745 & 54966.7094 & 1.821 & 1.618 & 0.403& A \\
  9944201 & 0.1618 & 0.1219 & 0.0 & 89.97 & 0.418 & 0.721523 & 54964.70882 & 0.524 & 0.064 & 0.2& A \\
  9945280 & 0.1708 & 0.1298 & 0.7101 & 89.72 & 0.001 & 1.3034599 & 54964.75158 & 0.824 & 0.107 & 0.277& A \\
  10014830 & 0.6542 & 0.7606 & 0.2601 & 89.99 & 0.007 & 3.0305267 & 54967.12282 & 3.621 & 2.754 & 0.605& A \\
  10129482 & 0.3567 & 0.7771 & 0.2093 & 77.47 & 0.001 & 0.8462906 & 54965.19403 & 0.829 & 0.644 & 0.316& B \\
  10189523 & 0.2051 & 0.2019 & 0.4387 & 83.64 & 0.002 & 1.0139159 & 54965.41787 & 0.823 & 0.166 & 0.204& A \\
  10257903 & 0.356 & 0.2697 & 0.6152 & 78.93 & 0.001 & 0.8585697 & 54965.0282 & 1.264 & 0.341 & 0.507& A \\
  10264202 & 0.3841 & 0.8712 & 0.6567 & 74.38 & 0.001 & 1.0351478 & 54965.55131 & 0.969 & 0.844 & 0.313& B \\
  10346522 & 0.5874 & 0.7852 & 0.186 & 85.85 & 0.004 & 3.989142 & 54965.57077 & 3.967 & 3.115 & 0.578& A \\
  10363300 & 0.3613 & 0.3173 & 0.2037 & 82.46 & 0.0 & 0.9349261 & 54965.17822 & 1.186 & 0.376 & 0.533& B \\
  10728219 & 0.1756 & 1.3494 & 0.0819 & 84.2 & 0.0 & 3.3718077 & 54971.24536 & 0.861 & 1.161 & 0.147& B \\
  11134079 & 0.2322 & 0.3485 & 0.1061 & 85.8 & 0.001 & 1.260566 & 54965.05979 & 0.994 & 0.346 & 0.312& A \\
  11147276 & 0.1702 & 0.4354 & 0.5664 & 84.29 & 0.0 & 3.1330583 & 54967.10369 & 1.228 & 0.535 & 0.183& B \\
  11198068 & 0.7353 & 0.6146 & 0.4375 & 76.89 & 0.011 & 0.4001746 & 54964.96665 & 1.206 & 0.741 & - & B \\
  11228612 & 0.1768 & 0.6333 & 0.2331 & 87.18 & 0.0 & 2.9804799 & 54967.40179 & 1.123 & 0.711 & 0.179& B \\
  11457191 & 0.3698 & 0.8381 & 0.9968 & 72.21 & 0.003 & 2.2983953 & 54965.96824 & 1.728 & 1.448 & 0.432& A \\
  11616200 & 0.2939 & 0.3987 & 0.1466 & 85.51 & 0.002 & 1.718649 & 54954.89948 & 1.467 & 0.585 & 0.375& B \\
  12004834 & 0.5799 & 0.5182 & 0.7619 & 69.25 & 0.001 & 0.2623168 & 55002.04157 & 0.537 & 0.278 & 0.57& B \\
  12022517 & 0.1955 & 0.6191 & 0.3319 & 81.29 & 0.002 & 3.4427135 & 55003.26138 & 1.243 & 0.769 & 0.2& B \\
  12023089 & 0.3376 & 0.9024 & 0.6898 & 72.88 & 0.0 & 0.6234413 & 55001.77468 & 0.608 & 0.548 & - & B \\
  12109575 & 0.3714 & 0.2392 & 0.4915 & 77.72 & 0.025 & 0.5316547 & 54953.62134 & 0.896 & 0.214 & - & A \\
  12109845 & 0.3317 & 0.4964 & 0.6406 & 84.09 & 0.002 & 0.8659588 & 55000.55844 & 0.941 & 0.467 & - & B \\
  12418662 & 0.257 & 0.1898 & 1.1052 & 81.28 & 0.004 & 2.751574 & 54967.30921 & 2.078 & 0.394 & 0.255& A \\
  12418816 & 0.2525 & 1.3472 & 0.7917 & 87.15 & 0.0 & 1.5218703 & 54954.74364 & 0.637 & 0.858 & 0.242& B \\
  12553806 & 0.6169 & 1.0979 & 0.3518 & 76.2 & 0.001 & 0.4631237 & 55000.04052 & 0.828 & 0.91 & - & A \\

\hline\hline
 
\end{tabular}
\label{Tab:param2}
\end{table*}


\bsp	
\label{lastpage}
\end{document}